\documentclass[twocolumn]{aastex631}
\usepackage{amsmath}
\usepackage{color}
\usepackage{graphicx,subfigure}

\graphicspath{{Fig/}}

\begin{document}

\title{Dynamical Consequence of Shadows Cast to the Outer Protoplanetary Disks: I. Two-dimensional Simulations}

% \correspondingauthor{Xue-Ning Bai}
% \email{xbai@tsinghua.edu.cn}

\author[0000-0001-5567-0309]{Zehao Su}
\affiliation{School of Physics and Astronomy, Beijing Normal University, Beijing 100875, China; suzh22@mail.bnu.edu.cn}
\affiliation{Institute for Frontier in Astronomy and Astrophysics, Beijing Normal University, Beijing 102206, China;}

\author[0000-0001-6906-9549]{Xue-Ning Bai}
\affiliation{Institute for Advanced Study, Tsinghua University, Beijing 100084, China; xbai@tsinghua.edu.cn}
\affiliation{Department of Astronomy, Tsinghua University, Beijing 100084, China}

\begin{abstract}
There has been increasing evidence of shadows from scattered light observations of outer protoplanetary disks (PPDs) cast from the (unresolved) disk inner region, while in the meantime these disks present substructures of various kinds in the submillimeter. As stellar irradiation is the primary heating source for the outer PPDs, the presence of such shadows thus suggest inhomogeneous heating of the outer disk in azimuth, leading to a ``thermal forcing" with dynamical consequences. We conduct a suite of idealized 2D disk simulations of the outer disk with azimuthally-varying cooling prescription to mimic the effect of shadows, generally assuming the shadow is static or slowly-rotating. The linear response to such shadows is two-armed spirals with the same pattern speed as the shadow. Towards the nonlinear regime, we find that shadows can potentially lead to the formation of a variety of types of substructures including rings, spirals and crescents, depending on viscosity, cooling time, etc. We have conducted systematic and statistical characterization of the simulation suite, and as thermal forcing from the shadow strengthens, the dominant form of shadow-induced disk substructures change from spirals to rings, and eventually to crescents/vortices. Our results highlight the importance of properly modeling the dynamical impact of inhomogeneous stellar irradiation, while call for more detailed modeling incorporating more realistic disk physics.
\end{abstract}


\section{Introduction} \label{Introduction}
Thanks to the advent of new observational facilities and instruments to conduct spatially
resolved observations of protoplanetary disks (PPDs), it has now been well established that disk
substructures are ubiquitous \cite[e.g.][]{Bae+PPVII}. In the millimeter/sub-millimeter wavelengths,
the Atacama Large Millimeter Array (ALMA) has revealed the richness of disk substructures that are primarily in the form of rings and gaps in addition to other features such as spirals and crescents at different radii \cite[e.g.][]{ALMA_Partnership2015,Monnier2017,Avenhaus2018,Isella2018,Huang2018b,Gratton2019,Andrews2020ARAA}. 
These observations reflect the thermal emission from mm-sized dusts around the disk midplane,
which are biased tracers of the gas density profiles due to finite aerodynamic coupling between
gas and dust.
In the optical and near infrared (NIR), high-contrast imaging with extreme adaptive optics
(e.g., VLT/SHERE, VLT/CRIRES, GPI) reveal even richer and more complex features \cite[e.g.][]{Benisty2015,Pinilla2015,Pohl2017a,Van_Boekel2017,Garufi2018,Benisty2022_PPVII}. Emission at optical/NIR mainly result from the starlight scattered by micron-sized dust (better coupled to the gas) suspended in the disk, which are better tracers of the disk surface layers. As a result, features seen in scattered light do not necessarily have direct correspondence to substructures recognized by ALMA \cite[e.g.][]{van-der-Marel2016,Uyama2018,Laura2018,Muro-Arena2018}.

At least partly contributing to the complexity in features seen in scattered light is the presence of shadows, typically defined as low-intensity regions that are confined to specific azimuthal angles \citep{Benisty2022_PPVII}. They must be cast from the (unresolved) disk inner region, and can be mainly classified into two types: broad extended shadows in azimuthal directions \cite[e.g.][]{Muro-Arena2020} and narrow shadow lanes with only a few degrees \cite[e.g.][]{Ginski2021}.
Considerable effort has been devoted to understanding the origin of shadows because the morphology and temporal variation of shadows can provide indirect information about the
disk's inner regions. The most common case for shadow casting is the presence of a misaligned/warped inner disk.
%Shadows in several sources are likely formed by this mechanism. 
For instance, TW Hya shows a moving shadow pattern that could suggests a precessing inner disk \citep{Debes2017}, shadows in HD 143006 can be reproduced using a $30^{\circ}$ misaligned inner disk \citep{Benisty2018}, fast time variations of shadows in RX J1604.3-2130A may come from dust very close to the central star in an irregular misaligned inner disk \citep{Pinilla2015}, narrow shadow lanes in SU Aur that possibly suggest misalignment caused by late-time interactions with infalling materials \citep{Kuffmeier2021}, shadows in HD 139614 can be explained by the combination of a misaligned inner ring and disk \citep{Muro-Arena2020}, and the flux ratio switches sides in  brightest nebula of IRAS40302 can be achieved by applying a tilted inner disk \citep{Villenave2023}. Even for disks with nearly aligned inner regions, subtle shadowing effects can still be recognized \citep{Monnier2017}. In addition to the misalignment of the inner disk regions, variations in the scale height of the inner disk atmosphere could also be responsible for generating shadows, such as in HD 163296 \citep{Rich2019,Rich2020,Varga2021,GRAVITY2021}.

Most effort aiming to understand disk shadows so far has focused on the modeling the (inner) disk morphology to explain the shadow features using radiative transfer calculations \cite[e.g.][]{Casassus2018,Benisty2018,Nealon2019,Muro-Arena2020}. On the other hand, we note that as stellar irradiation is the primary source of heating in the bulk of the (outer) PPDs, the presence of shadows must also give rise to dynamical consequences by ``thermal forcing": the disk gas experiences (quasi-)periodic cooling and heating as it enters and exits the shadow, which hardly settles to thermal equilibrium and constantly exerts modest or even strong pressure perturbations to the neighboring fluid. This effect was first explored in \cite{Montesinos2016}, who conducted 2D hydrodynamic simulations that take into account both stellar irradiation and periodic forcing of shadows with an opening angle of $28^{\circ}$ in the context of the transition disk HD 142527. They found that azimuthal pressure gradients generated by shadows can trigger $m=2$ spirals, which are enhanced by self-gravity and give rise to observable quasi-steady spiral signals.

In this work, motivated by the diversity of shadowing features seen in scattered light images and the case study by \cite{Montesinos2016} for the HD 142527 disk, we aim at a systematic exploration on the dynamical consequences of shadows cast onto outer PPDs. As an initial effort, we restrict ourselves to vertically-integrated systems by 2D hydrodynamic simulations. We follow the evolution of a passive, viscous gaseous disk with a thermal relaxation prescription towards a target temperature, which is set by stellar irradiation subject to shadowing. By exploring a large suite of simulations varying the viscosity, cooling time, and shadow geometry, we find that shadows can result in the formation of a wide variety of disk substructures, and perform a statistical analysis of all the substructures generated from our simulations.

This paper is organized as follows. We detail our simulation setup in Section \ref{Sec:Methods}, followed by a description representative features of shadow-driven disk substructures in Section \ref{Sec:Results}. We present statistical analysis of all our simulations and describe the substructure-forming process from linear to non-linear regimes in Section \ref{Sec:Statistics}. Finally,  we discuss the caveats and conclude in Section \ref{Sec:Conclusions}.

\section{Numerical Methods} \label{Sec:Methods}

\subsection{Simulation Setup} \label{subSec:Governing}

We solve the vertically integrated viscous hydrodynamic equations using the grid-based higher-order Godunov code ATHENA++ \cite{Stone2020} in cylindrical coordinates $(r,\varphi)$. The conservative form of control equations are:
\begin{equation} \label{Equ:continuity}
\frac{\partial \Sigma}{\partial t}+ \nabla \cdot (\Sigma\textbf{v})=0, 
\end{equation}

\begin{equation} \label{Equ:momentum}
\frac{\partial \Sigma\textbf{v}}{\partial t}+ \nabla \cdot (\Sigma\textbf{v}\textbf{v} + P\mathcal{I} + \mathcal{T}_{vis})=-\Sigma\nabla\Phi,
\end{equation}

\begin{equation} \label{Equ:energy}
    \frac{\partial E}{\partial t} + \nabla \cdot \left[(E + P + \mathcal{T}_{vis})\textbf{v}\right]=-\Sigma\textbf{v}\cdot\nabla\Phi + \Lambda,
\end{equation}
where $\Sigma$ is the gas surface density in the disk, $\textbf{v}$ is the gas velocity, $P=\Sigma T$ is the vertically integrated pressure where $T$ is the disk temperature, $\mathcal{I}$ is the identity tensor, $\Phi$ is the gravitational potential written as $\Phi = -GM_{*}/r$ where $M_{*}$ is mass of central star, $E$ is total energy density, and $\Lambda$ gives the cooling source terms.
The viscous stress tensor $\mathcal{T}_{visc}$ in momentum equation reads
\begin{equation} \label{Equ:possion}
\mathcal{T}_{vis} = -\Sigma\nu\left( \frac{\partial v_{i}}{\partial x_{j}} + \frac{\partial v_{j}}{\partial x_{i}} - \frac{2}{3}\frac{\partial v_{k}}{\partial x_{k}}\delta_{ij} \right),
\end{equation}
with $\nu$ being the kinematic viscosity.

The total energy density $E$ is given by the combination of kinetic energy and internal energy:
\begin{equation}
E = \frac{1}{2}\Sigma v^{2} + \frac{P}{\gamma-1},
\end{equation}
where $\gamma=7/5$ is the adiabatic index for molecular gas. Note that viscous heating is automatically included in the energy equation, although it is generally unimportant in the outer PPDs.
The gas temperature $T$ is associated with the isothermal sound speed as $T=c_s^2$,
which yields the disk scale height $H=c_s/\Omega_K$, where $\Omega_K=(GM_*/r^3)^{1/2}$
is the Keplerian angular frequency. The disk aspect ratio is then given by $h=H/r$.
With this, viscosity follows the standard $\alpha$ prescription \citep{Shakura1973}, $\nu=\alpha c_sH$. It is worth noticing that viscosity varies as the disk evolves. 

\begin{figure*}[htbp]
    \centering
    \includegraphics[width=0.9\textwidth]{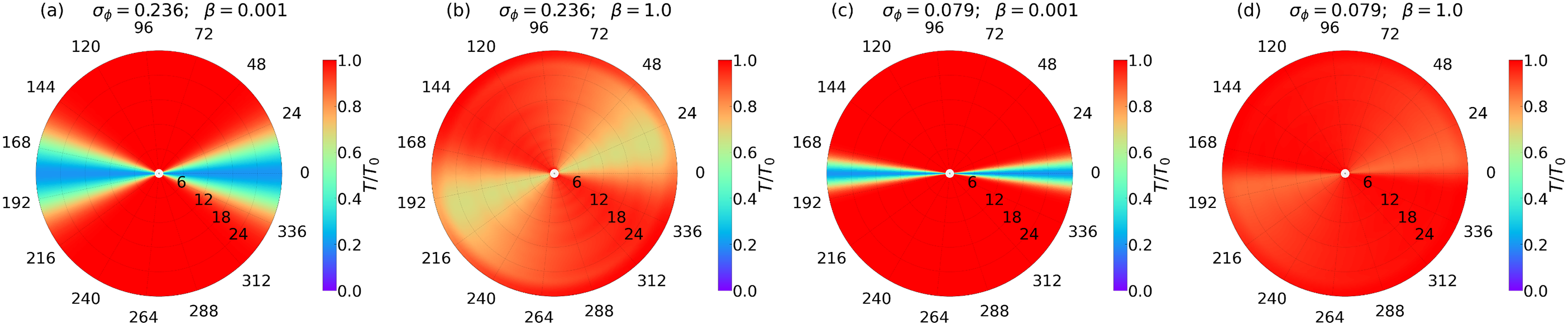}
    \caption{Illustration of our shadow morphology depicted by normalized temperature in disks with $\sigma_{\phi}=0.236$ and $0.079$ for fast ($\beta=0.001$) and slow ($\beta=1$) cooling processes. With fast cooling, the low-temperature regions caused by shadows largely reflect the target temperature we prescribe at the shadow location. With longer cooling timescale, these regions deviate from the prescribed shadow positions towards the leading side, and exhibit significantly weaker amplitude.}
    \label{Fig:temperature}
\end{figure*}

We choose the initial density profile to be power-law:
\begin{equation}
\Sigma_{\rm init} = \Sigma_{0}\left( \frac{r}{r_{0}}\right)^{d}.
\end{equation}
where $\Sigma_{0}$ is the density at reference radius $r_{0}$, which we take to be the radius of the inner boundary.
We specify the disk initial temperature as 
\begin{equation} \label{Equ:T_{init}}
T_{\rm init} \equiv c_{s_{0}}^{2}\left( \frac{r}{r_{0}}\right)^{p}=\left( h_{0}r_{0}\Omega_{K_{0}}\right)^{2}\left( \frac{r}{r_{0}}\right)^{p},
\end{equation}
where $c_{s_{0}}$, $h_{0}$, $\Omega_{K_{0}}$ are the value of isothermal sound speed, aspect ratio and Keplerian angular velocity at reference radius $r_{0}$.

The radial force balance leads to the initial rotation profile calculated by
\begin{equation}
v_{\varphi}(r)=\left[\left(p+d\right)c_{s}^{2}+\frac{GM_{*}}{r}\right]^{1/2}\ .
\end{equation}
With viscosity, the radial velocity is set by the accretion velocity given by
\begin{equation}
v_{r}(r)=-\frac{3}{2}\frac{\alpha c_{s}^{2}}{r\Omega_{K}}\ .
\end{equation}
To ensure steady-state accretion in the initial equilibrium (shadow not included) with constant
$\alpha$, the initial temperature and density profile should satisfy $d+p=-3/2$.

Our simulations are scale-free, adopting $GM_{*}=\Sigma_{0}=r_{0}=1$ in code units, with $h_{0}=0.1$. As a result, we have $\Omega_{K_{0}}=c_{s_{0}}=1$. The computational domain ranges from $r_{\rm in}=1$ to $r_{\rm out}=30$ in code units, to ensure sufficient dynamical range. We employ a logarithmic grid in radial direction and a uniform grid in azimuthal direction with $N_{r}\times N_{\varphi}=512\times512$, achieving a gird resolution of 15 cells per $H$ in $r$ while keeping cell aspect ratio $\Delta r\approx 0.5r\Delta\phi$.

\subsubsection[]{Shadow prescription}

In our simulations, we assume an obscuring structure present in disk inner region, which is outside of our simulation domain (inside the inner boundary). As an initial study, we tentatively take this obscuring structure as a slightly misaligned inner disk (inclination angle $\sim h$). In this case, near half of the outer disk (in azimuth) is illuminated from one hemisphere, and the opposite side being illuminated from the other hemisphere. In the vertically-integrated sense, the two sides of the ``pie-chart" are heated largely equally. It is the transition region, the disk can be largely blocked by the inner disk from both hemisphere, which is mostly affected by the shadow. The shadow then introduces a thermal forcing to the system, causing system's temperature to approach the target temperature $T_{\rm tar}$. For simplicity, we prescribe this target temperature by
\begin{equation}
T_{\rm tar} (r,\phi)= T_{\rm init}(r)\left( 1-\epsilon e^{-\frac{\phi^{2}}{2\sigma_{\phi}^{2}}}\right)\left( 1-\epsilon e^{-\frac{(\phi-\pi)^{2}}{2\sigma_{\phi}^{2}}}\right),
\end{equation}
where $\epsilon$ reflects the amplitude of the shadow and $\sigma_{\phi}$ characterizes the azimuthal width of the shadow. Although we have argued that two-sided shadows are the most basic case, we still examine the one-sided case in \ref{App:one shadow} for reference. Also, in most cases in this paper, except for the simulation mentioned in Section \ref{subSec:rotating shadow}, the shadow is static with pattern speed ($\Omega_{\rm shadow}$) being zero.

The final temperature structure depends on heating and cooling process, which is often modeled using the $\beta$ cooling approximation \citep{Gammie2001}. The cooling term is given by thermal relaxation towards the target temperature
\begin{equation}
\Lambda = -\frac{\Sigma}{\gamma-1}\times\frac{T-T_{\rm tar}}{t_{\rm cool}}\ ,
\end{equation}
where the cooling timescale is specified by the dimensionless parameter $\beta$:
\begin{equation}
t_{\rm cool} = \beta\Omega_{K}^{-1}\ .
\end{equation}
It describes the disk's thermodynamic timescale, which can range from $\sim10^{-3}$ (approaching the isothermal limit) to at least $\sim10$ (approaching the adiabatic limit) in our simulations. Figure \ref{Fig:temperature} shows the expected temperature structure for four representative shadow prescriptions, with different shadow widths ($15^{\circ}$ and $45^{\circ}$) and cooling times, calculated by following fluid elements undergoing heating and cooling on circular orbits. We have fixed shadow amplitude of $\epsilon=0.8$. With fast cooling ($\beta=0.001$), we see that the shadow aligns with its expected position, and the temperature at its center is approximately $0.2T_{0}$ as desired. When cooling is inefficient, the observed shadow center on the leading side from its expected location, and the lowest temperature becomes well above $0.2T_{0}$.

It should be noted that our shadow and cooling prescriptions are highly simplified and are not necessarily always physical (for instance, a flat disk with $p=-1$ would not be irradiated). We emphasize that the goal of this work is not to precisely model any particular system, but to explore the general phenomenology in a qualitative manner.

\subsubsection{Boundary Conditions} \label{subsubSec:boundary}

We use modified outflow boundary conditions, where hydrodynamic variables are copied from the last grid zone assuming $\Sigma\propto r^{d}$, $P\propto r^{d+p}$, $v_{\phi}\propto r^{-1/2}$, with $v_{r}$ unchanged---except we set $v_{r}=0$ in case of inflow. To further dampen unphysical waves, we adopt wave-killing functions in the form described by \cite{Val-Borro2006}:
\begin{equation}
\frac{dx}{dt} = -\frac{x-x_{0}}{\tau_{\rm damp}}R(r)
\end{equation}
where $x$ represents any fluid quantities (e.g. $\Sigma$, $\textbf{v}$, etc.). The damping timescale $\tau_{\rm damp}$ is defined as $\tau_{\rm damp}=\eta\Omega_{K}^{-1}$, where $\eta$ is the damping rate and is set to 1 for all of our simulations. The function $R(r)$ is a parabolic function expressed as:
\begin{equation}
R(r) = \left(\frac{r-r_{\rm damp}}{L_{\rm damp}}\right)^{2},\ {\rm for} \left|r-r_{\rm damp}\right|<L_{\rm damp}\ ,
\end{equation}
where $r_{\rm damp}$ is the boundary of the damping area which we take to be $2.08$ and $26.57$ in inner and outer part of our computation domain, respectively, and $L_{\rm damp}$ is the length of wave killing zone.

\subsection{Simulation Runs} \label{subSec:parameter}

In order to comprehensively investigate the dynamical effects of shadows in PPDs, we conducted a wide parameter scan. Five main parameters are included in our simulations: dimensionless cooling timescale $\beta$ ranging from $10^{-3}$ to $10$, viscosity coefficient $\alpha$ ranging from $0$ to $10^{-2}$, shadow amplitude coefficient $\epsilon=0.5$ and $\epsilon=0.8$, shadow width $\sigma_{\phi}=0.236$ and $\sigma_{\phi}=0.079$, and the temperature slope $p=-1$ (flat case) and $p=-0.5$ (flared case). In the viscous simulations, they translate to density gradient $d=-0.5$ and $d=-1$ to ensure steady state accretion. In most simulations, the shadows do not rotate, and we fix the disk aspect ratio $h_0=0.1$ at $r=1$, thus $h=0.1$ is constant in most $p=-1$ (flat) cases. All of these simulations will be discussed in Section \ref{Sec:Statistics}. In Sections \ref{subSec:rotating shadow} and \ref{subSec:different aspect ratio}, we also briefly explore simulations with rotating shadows and vary $h_0$ from $0.05$ to $0.15$.

To further comment on our choice of parameters, we first note that in outer disk conditions, we generally expect $\beta\lesssim1$ \cite[e.g.][]{Lin&Youdin2015,Pfeil&Klahr2019}, though the finite thermal coupling between dust and gas may significantly enhance the effective $\beta$ \cite[e.g.][]{Bae2021ApJ...912...56B}. In inviscid simulations, we further examine the influence of the density profile ($d=-0.5$ and $d=-1$, which affects thermal forcing) since this parameter is no longer free when viscosity is included (dependent on $p$). Note that in the new paradigm of wind-driven accretion, the disk is more laminar and the surface density profile can be more arbitrary \cite[e.g.][]{Bai2016ApJ...821...80B,Suzuki2016A&A...596A..74S,Tabone2022MNRAS.512.2290T}. Although we do not incorporate wind-driven accretion, this exploration also serves the purpose to partly mimic ``windy" disk conditions. On the choice of shadow amplitudes, note that given the $T^4$ dependence, the two choices correspond to the shadowed region receiving about $0.2^4\approx0.002$ and $0.5^4\approx0.06$ of the stellar irradiation compared to the non-shadowed regions.

In all of our simulations, the total run time is chosen to be $T=20000P_0$, where $P_0=2\pi/\Omega_{K_{0}}$ is the orbital period at the inner boundary. This is significantly longer than the timescales for substructure formation, which we find to be within $5000P_0$ for most cases. In only a few cases (especially $\beta\sim 10, \alpha\sim 0$), even on the timescale of $20000P_0$, we cannot unambiguously identify the dominant form of disk substructure. However, we can infer their evolution trend from a statistical point of view. 

To facilitate comparison of the various simulations discussed in the following sections, we provide a list of all our runs and their parameters in Table \ref{Tab:table-1}. Our naming convention is structured as follows. We use ``L" for runs in the linear regime ($\epsilon=0.001$) and ``NL" for runs in the nonlinear regime ($\epsilon=0.5, 0.8$). The labels ``hs", ``hm" and ``hl" indicate runs with $h_0=0.05$, $h_0=0.1$, and $h_0=0.15$, respectively, with $h_0=0.1$ as fiducial. To specify the dominant substructure, we use ``S" for spiral-dominant, ``R" for ring-dominant, and ``V" for vortex-dominant. Shadow precession speeds are denoted as ``NR" for nonrotating (fiducial), ``FR" for fast rotating, ``MR" for moderately rotating, and ``SR" for slow rotating. For simulations dedicated to parameter searches discussed in Section 4, we use the label ``S-h-all," as we do not discuss individual runs for these simulations.

%%%%%%%%%%%%%%%%%%%%%%%%%%%%%%%%%%%%%%%%%%Table1-Part1%%%%%%%%%%%%%%%%%%%%%%%%%%%%%%%%%%%%%%%%%%%%%%%%
\setlength{\tabcolsep}{3mm}
\begin{deluxetable*}{lccccccccc}[!htb]
\tabletypesize{\scriptsize}
\tablecolumns{8}
% \tablewidth{0.8\textwidth}
%\tabletypesize{\tiny}
\tablecaption{Summary of All Highlighted Simulations.$^{1}$ \label{Tab:table-1}}
\tablenum{1}
\tablehead{
\colhead{Run} & \colhead{$\sigma_{\phi}$} & \colhead{$\epsilon$} & \colhead{$\alpha$} & \colhead{$\beta$} & \colhead{p} & \colhead{$h_{0}$} & \colhead{$\Omega_{\rm shadow}$}}

\startdata
\hline
\multicolumn{8}{l}{Representative runs (Section \ref{Sec:Results})}\\
\hline
NL-hm-S-NR  & 0.236    & 0.8   & $10^{-3}$       & 10     & -1.0   & 0.1          & 0   & \\
NL-hm-R-NR  & 0.236    & 0.5   & $10^{-4}$       & 1     & -1.0   & 0.1          & 0   & \\
NL-hm-V-NR  & 0.236   & 0.5 & $0$      & $10^{-3}$    & -1.0   & 0.1          & 0   & \\
\hline
\multicolumn{8}{l}{Statistical runs (Section \ref{Sec:Statistics})}\\
\hline
S-h-all$^{2}$  & $(0.236, 0.079)$     & $(0.5, 0.8)$    & $(0, 10^{-4}, 10^{-3}, 10^{-2})$      & $(10^{-3}, 10^{-2}, 10^{-1}, 1, 10)$    & $(-1.0, -0.5)$   & 0.1          & 0   & \\
\hline
\multicolumn{8}{l}{Linear run (Section \ref{subSec:linear regime})}\\
\hline
L-hm-S-NR  & 0.236      & 0.001  & 0       & $10^{-3}$   & -1.0   & 0.1          & 0        & \\
\hline
\multicolumn{8}{l}{Rotating shadow runs (Section \ref{subSec:rotating shadow})}\\
\hline
L-hm-S-FR$^{3}$  & 0.236   & 0.001   & 0  & $10^{-3}$   & -1.0   & 0.1      & $\Omega_{0}$   & \\
L-hm-S-MR  & 0.236   & 0.001   & 0       & $10^{-3}$     & -1.0   & 0.1          & $0.03\Omega_{0}$   & \\
L-hm-S-SR  & 0.236    & 0.001 & 0       & $10^{-3}$     & -1.0   & 0.1          & $0.003\Omega_{0}$  & \\
\hline
\multicolumn{8}{l}{Aspect ratio test runs (Section \ref{subSec:different aspect ratio})}\\
\hline
NL-hs-S-NR  & 0.236    & 0.8   & $10^{-3}$       & 10     & -1.0   & 0.05          & 0   & \\
NL-hs-R-NR  & 0.236    & 0.5   & $10^{-4}$       & 1     & -1.0   & 0.05          & 0   & \\
NL-hs-V-NR  & 0.236   & 0.5 & 0      & $10^{-3}$    & -1.0   & 0.05          & 0   & \\
NL-hl-S-NR  & 0.236    & 0.8   & $10^{-3}$       & 10     & -1.0   & 0.15          & 0   & \\
NL-hl-R-NR  & 0.236    & 0.5   & $10^{-4}$       & 1     & -1.0   & 0.15          & 0   & \\
NL-hl-V-NR  & 0.236   & 0.5 & $0$      & $10^{-3}$    & -1.0   & 0.15          & 0   & \\
\enddata
\tablenotetext{1}{Simulations mentioned in Appendix \ref{App:transition state} and Appendix \ref{App:one shadow} are not included.}
\tablenotetext{2}{The unified name for parameter sruvey simulations, with parameters being all combinations of those listed in this row, totaling 160 simulations.}
\tablenotetext{3}{The resolution of this run is set to $N=2048$.}
\tablecomments{$\sigma_{\phi}$: shadow width parameter; $\epsilon$: shadow amplitude parameter; $\alpha$: viscosity parameter; $\beta$: cooling rate parameter; p: temperature slope; $h_{0}$: disk aspect ratio at inner boundary; $\Omega_{\rm shadow}$: shadow procession angular frequency; $\Omega_{0}$: Keplerain angular velocity at $r=1$. All runs, except for run L-hm-S-FR, use a resolution of $N=512$.}
\end{deluxetable*}

\subsection{Diagnostics of Substructures} \label{subSec:diagnostics}

As we will demonstrate, our simulations generate a variety of substructures of all types. In this section, we provide the diagnostics we employ to identify and characterize these substructures. To minimize the influence of the boundaries and wave-killing, we restrict the analysis domain to be $r\in[3,21]$.

Vortices exhibit as anti-cyclonic flows with pressure maxima at the center that can potentially be strong dust traps. They are identified as regions with negative vorticity, which is defined as $\nabla\times\delta\textbf{v}$ with $\delta\textbf{v}$ being the difference between current fluid velocity and background fluid velocity. We quantify individual vortices based on their mean vorticity (normalized by background Keplerian angular velocity), density contrast, spacing, and aspect ratio. In doing so, we first choose the vortex boundary to be where the density is $10\%$ of the density at the vortex center after subtracting background, while ensuring that the vorticity remains below zero. This is motivated from the analytical work of \cite{Lyra2013ApJ...775...17L} while being robust to the influence of density waves.
In our simulations, vortices are constantly generated and destroyed; only the largest vortices are chosen (usually can survive for at least 100 local orbits). We measure the density contrast by comparing the average density in vortex with the average density at the same radius. The spacing of vortices is calculated by the radial distances between neighboring vortices, which are normalized by the local scale height at the midpoint radius between the two vortices. As vortices can be highly time variable, all quantities are calculated and averaged over several snapshots (see in Section \ref{Sec:Statistics}).

In ring-forming disks, we measure the density contrast, width, spacing and eccentricity of the rings. We identify the rings by first fitting the background density as a power law, and consider peaks/troughs above/below the fitted profile as ring peaks/gap centers. The boundaries of the rings are identified as the radius at the midpoint between peak and valley densities, with ring width being the distance between the two boundaries for each ring. Density contrast is calculated by comparing the density between peaks and boundaries. The final ring width is obtained by averaging the widths of all identified rings, and each ring width is normalized to the local scale height of the disk. Ring spacing is measured as the radial distances between the boundaries of two neighboring rings, normalized in a way similar to that for vortices, and averaged over several snapshots. In the above, we have treated the rings as axsymmetric by working with 1D profiles, whereas in practice we have found that the rings can be eccentric. For identified rings, we further track the maximum density in 2D data and measure their eccentricity by fitting an ellipse. Incomplete rings at the boundary of the analysis domain are excluded from the statistics.

For spirals, we quantify their density contrast, number of spiral arms, pattern speed and pitch angle. The density contrast is obtained by comparing density of the spiral spine and the fitted background density at same radii. In our simulations, we obtain the spiral phase at each radius using Fourier decomposition and the pitch angle is obtained by fitting the phase angle with a logarithmic function $\varphi=m(\tan\alpha_{p})^{-1}\ln r+\phi_0$, where $\alpha_{p}$ is the pitch angle and $m$ is number of spiral arms.
The constant $\phi_0$ is further employed to measure the pattern speed of the spirals.

\section{Representative Results}  \label{Sec:Results}

In this section, we present representative outcomes of shadow-driven substructures at fixed disk aspect ratio $h$ before giving more comprehensive statistical results. The three representative runs, denoted as``NL-hm-S-NR," ``NL-hm-R-NR," and ``NL-hm-V-NR," can be found in Table \ref{Tab:table-1}. We show snapshots of major fluid quantities of interest (i.e. $\Sigma, T, \textbf{v}_{r}, \textbf{v}_{\phi}, \nabla\times\textbf{v}$) from our simulations, and discuss the results below.

\subsection{Spirals} \label{subSec:spiral}
\begin{figure*}[htbp]
    \centering
    \includegraphics[width=0.8\textwidth]{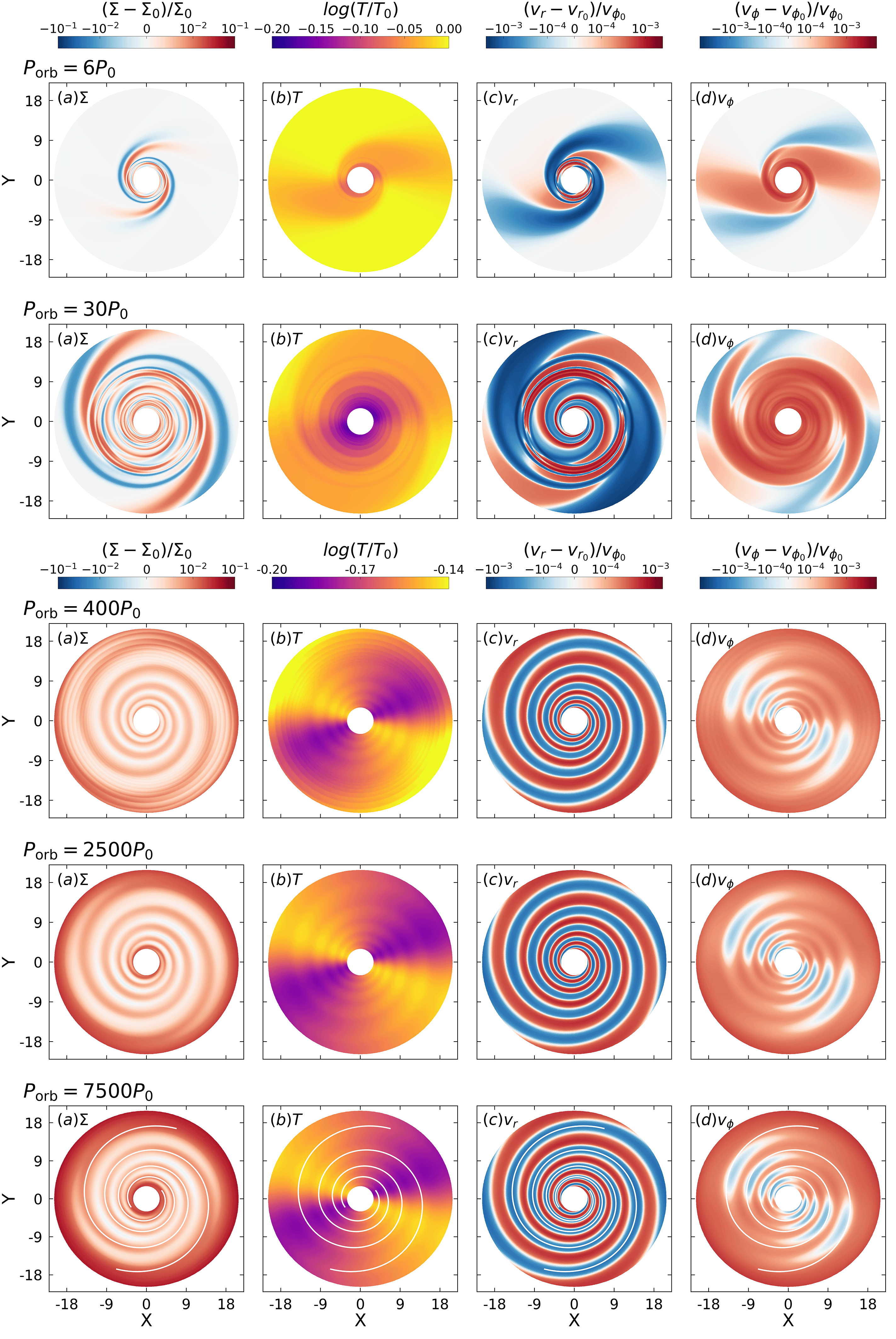}
    \caption{Density (a), temperature (b), radial velocity (c) and azimuthal velocity (d) evolution in spiral-forming disk ($\sigma_{\phi}=0.236, \epsilon=0.8, \alpha=10^{-3}, \beta=10, p=-1.0$). All quantities except for $v_{r}$ are normalized by their initial values, and $v_{r}$ is normalized by initial $v_{\phi}$. The color range for temperature differs between the formation process of the early stage (first and second rows) and the relative quasi-steady state (third to fifth rows). The white solid lines represent the fitted spirals based on density contrast.}
    \label{Fig:spiral}
\end{figure*}

Spirals typically form in disks where the shadow is relatively weak, such as those characterized by slow cooling or weak shadow amplitude. We choose spirals formed in a disk with the following parameters as an example: $\sigma_{\phi}=0.236, \epsilon=0.8, \alpha=10^{-3}, \beta=10, p=-1.0$ (run NL-hm-S-NR). As depicted in Figure \ref{Fig:spiral}, spirals form relatively quickly (first row in Figure \ref{Fig:spiral}), typically within approximately 20 local orbits, and once formed, they remain highly stable\footnote{The growth in density perturbations observed in the last two rows of Figure \ref{Fig:spiral} is primarily due to the combined effects of strong viscous heating and the influence of wave damping zones.}. These spirals are clearly density waves, showing spiral patterns in all diagnostic physical quantities. The spiral patterns are stationary (i.e., pattern speed is zero), which is related to the fact that our shadow patterns have zero angular velocity. Further discussions regarding the relationship between the properties of the spirals and the other two substructures will be provided in Section \ref{subSec:nonlinear regime}. In addition, by examining the second column of Figure \ref{Fig:spiral}, we see that with inefficient cooling, the overall temperature is systematically cooler than the initial temperature by $\sim15\%$. The azimuthal varies smoothly through the shadowing regions, with a maximum temperature variation of about $4\%$.

\subsection{Rings} \label{subSec:ring}
\begin{figure*}[htbp]
    \centering
    \includegraphics[width=0.8\textwidth]{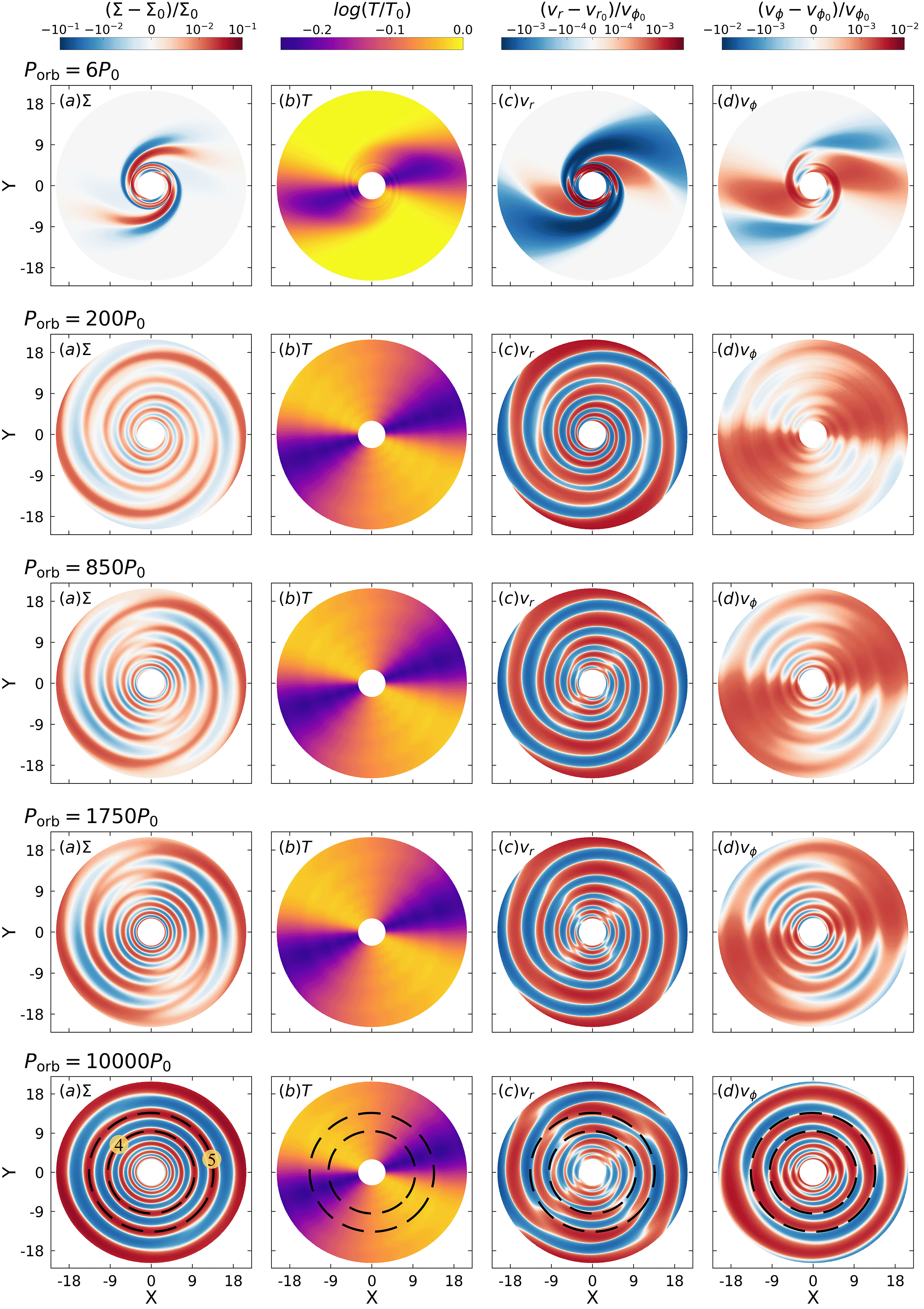}
    \caption{Density (a), temperature (b), radial velocity (c) and azimuthal velocity (d) evolution in ring-forming disk ($\sigma_{\phi}=0.236, \epsilon=0.5, \alpha=10^{-4}, \beta=1, p=-1.0$). All quantities except for $v_{r}$ are normalized by their initial values, and $v_{r}$ is normalized by initial $v_{\phi}$. The black dashed lines represent the fourth and fifth fitted rings in this disk.}
    \label{Fig:ring}
\end{figure*}

The conditions for ring formation generally require either slow cooling or a combination of moderate viscosity and shadow amplitude (for more detailed information, see Section \ref{Sec:Statistics}). In Figure \ref{Fig:ring}, we adopt parameters $\sigma_{\phi}=0.236, \epsilon=0.5, \alpha=10^{-4}, \beta=1, p=-1.0$ (run NL-hm-R-NR) to illustrate the typical formation process and properties of ring structures.

--{\it Formation}. The formation of rings begins with the presence of two-arm spirals following a transient period (as seen in the first to third rows of Figure \ref{Fig:ring}). The spirals appear only marginally stable, which later break apart and reconnect to form concentric rings in surface density (as shown in the fourth and fifth rows of Figure \ref{Fig:ring}), which takes a relatively long time of $\sim 100$ local orbits. On the other hand, the spiral patterns remain in the velocity structure even after ring formation, although they are distorted (as opposed to the spirals discussed in Section \ref{subSec:spiral} and could become a distorted ring patterns in some cases). 

--{\it Evolution and main properties}.  
Once formed, the amplitudes of the rings continue to increase slowly, reaching a steady state over a few hundred local orbits, where the gas density in rings are about $10\%$ higher than the background. However, the density within one ring at quasi-steady state is unevenly distributed, with surface density near the broken/reconnection location being smaller, which will be further discussed in Section \ref{subSec:nonlinear regime} and Appendix \ref{App:transition state}. The typical ring width is approximately twice the local scale height, and the spacing is regular (about $4H$ between peaks of two neighbouring rings) across the disk (further discussed in Section \ref{Sec:Statistics}). 
We find the rings to be eccentric (but centered on the star), with the eccentricity measured to be $e\sim0.12$. As can be inferred from the third and fourth columns in Figure \ref{Fig:ring}, the ratio $v_{r}/v_{\phi}$ is approximately $10^{-3}\ll e$, suggesting that these rings do not directly correspond to the gas moving in eccentric orbit. Also, the eccentric rings do not precess, analogous to spiral patterns that remain stationary, thus coroborating the fact that the rings emerge as the aftermath of spiral patterns. With moderate cooling, the azimuthal temperature contrast reaches 15$\%$ and may cause azimuthal brightness variations in observed rings, although we caution for our highly simplified thermodynamic treatment (further discussed in Section \ref{subSec:Observation}).

\subsection{Vortices and Crescents} \label{subSec:vortex}
\begin{figure*}[htbp]
    \centering
    \includegraphics[width=0.85\textwidth]{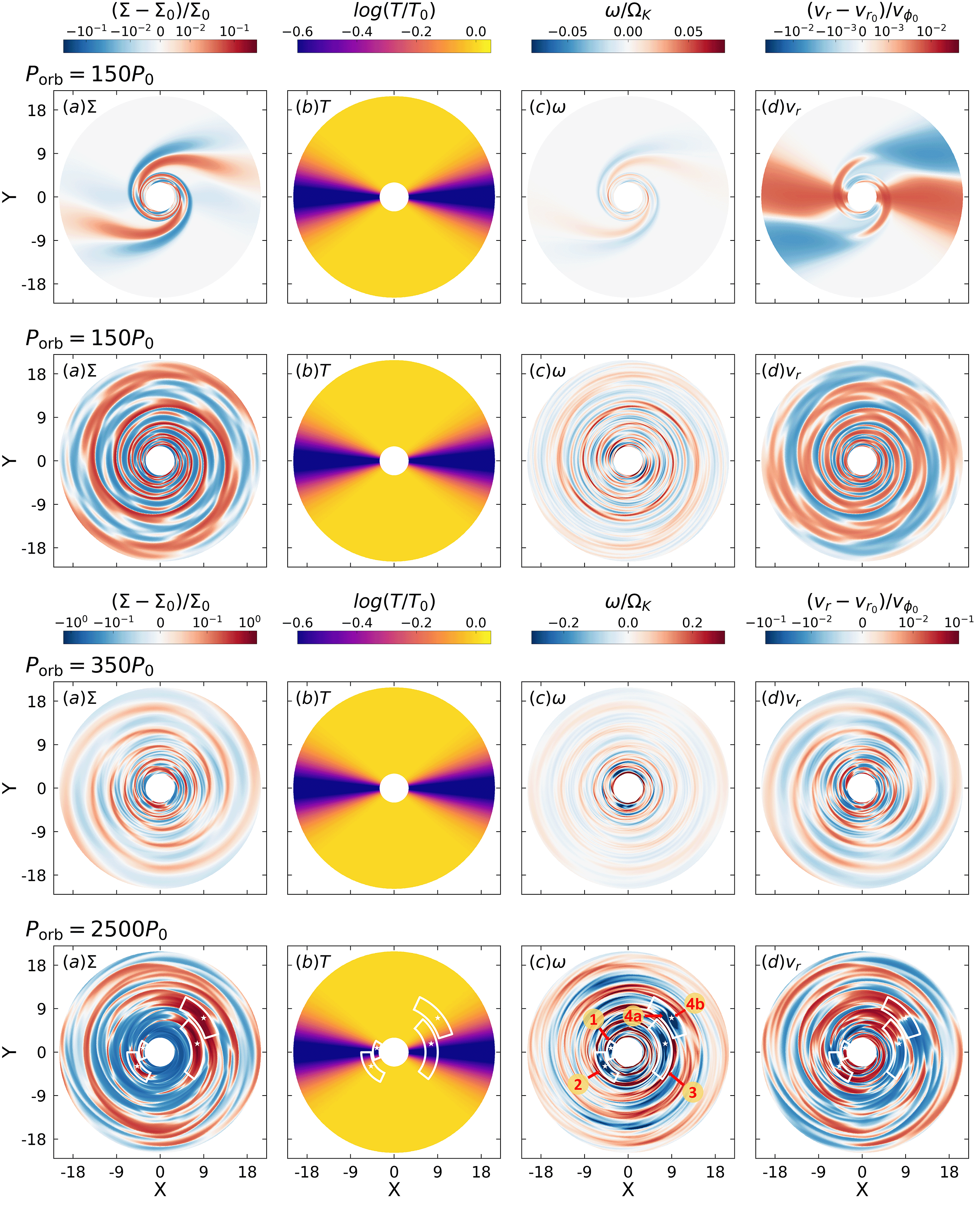}
    \caption{Density (a), temperature (b), vorticity (c) and radial velocity (d) evolution in vortex-forming disk ($\sigma_{\phi}=0.236, \epsilon=0.5, \alpha=0, \beta=0.001, p=-1.0$). The color ranges for density, vorticity and radial velocity differ between the early spiral formation stage (first and second rows) and the vortex-dominant stage (third and fourth rows). The simulation duration increases sequentially from top to bottom.  The white frame delineates the boundary of each identified vortex, and there are no significant structures inside the white frame in the temperature plot (b) (compare with density (a), vorticity (c) and radial velocity (d)) due to rapid cooling. A white star within each frame indicates the center of the vortex. The red numbers highlight the selected vortices. 
    }
    \label{Fig:vortex}
\end{figure*}

Crescents can be described as rings that exhibit an azimuthal variation in intensity \citep{Bae+PPVII}. Physically, the crescents discussed in this paper are all induced by vortices, thus we use ``vortices" and ``crescents" interchangeably.
Figure \ref{Fig:vortex} shows an example of shadow-driven formation of vortices/crescents. This usually occurs with strong shadow amplitude and rapid cooling, thus strong thermal forcing, and we adopt $\epsilon=0.5, \beta=0.001$ in this example (run NL-hm-V-NR).

--{\it Formation}. 
With rapid cooling, the disk temperature almost instantly relaxes to the target temperature both within and outside of the shadow region, resulting in a 50$\%$ variation in azimuthal temperature given our setup. This leaves two symmetric low-pressure regions that form quickly at the shadow locations. In the initial stages (first and second rows in Figure \ref{Fig:vortex}), it leads to the appearance of spiral features in surface density. With strong thermal forcing that constantly perturbing the disk, the system subsequently becomes more chaotic (third row in Figure \ref{Fig:vortex}) where the velocity field undergoes significant alterations. Although the physical process is not entirely clear, vortex/crescent formation ensues, as identified in fourth row of Figure \ref{Fig:vortex}. Selected vortices and crescents are marked by white frames in Figure \ref{Fig:vortex}.

--{\it Evolution and main properties}. Shadow-driven vortices are all anti-cyclonic in nature, which can be observed either from the negative vorticity (the 3rd column of Figure \ref{Fig:vortex}) or from the change in the sign of radial velocity across the vortex center (changing from negative to positive when viewed along the direction of rotation (counterclockwise), as shown in the 4th column of Figure \ref{Fig:vortex}).
We observe that vortices started small and are continuously generated. They merge to form larger ones under the influence of differential rotation within approximately 60 local orbits, ultimately manifesting as relatively large crescent-shaped structures.
In Figure \ref{Fig:vortex}, vortices labeled as 4a and 4b are undergoing a merger into one single vortex.  We find that these patterns largely corotate with the gas, as expected, and their azimuthal locations are found to be largely random, with no preference to stay in or out of the shadows. 
The disk gas remains turbulent and chaotic throughout the evolution due to strong perturbations from thermal forcing. Velocity deviations from local Keplerian inside the vortex region are around $0.5c_{s}$. Additionally, the local level of turbulence, measured in terms of root mean square (rms) velocity fluctuations averaged in azimuth, is approximately $10\%$ of the local sound speed. The typical aspect ratio of the vortices/crescents is about 6, with their density contrast being 1.4. The normalized vorticity in this case is 0.2. Despite of modest to strong level of turbulence, the large vortices are relatively long-lived, with typical lifetime of at least 300 local orbits.
 
\section{Statistics of Substructures}  \label{Sec:Statistics}

To gain deeper insights into the dynamical consequences of shadows, we conducted a comprehensive exploration of parameter space. We performed a total of 160 simulations (run S-h-all), encompassing a wide combination of parameters. Most results show similarities with one of the aforementioned three representative cases. We thus primarily summarize the outcomes in a statistical manner.

For simulations that exhibit the formation of rings and spirals, we only measure their properties at the end of the simulations when the system has already reached a steady state. For simulations with vortex/crescent formation which are generically chaotic, we select four specific snapshots, denoted as $P_{\rm orb1}=5000P_0$, $P_{\rm orb2}=10000P_0$, $P_{\rm orb3}=15000P_0$, and $P_{\rm orb4}=20000P_0$. The statistical values for vorticity, density contrast, spacing, and aspect ratio of the vortices are calculated by averaging the results at these snapshots. 

\begin{figure*}[htbp]
    \centering
    \includegraphics[width=0.9\textwidth]{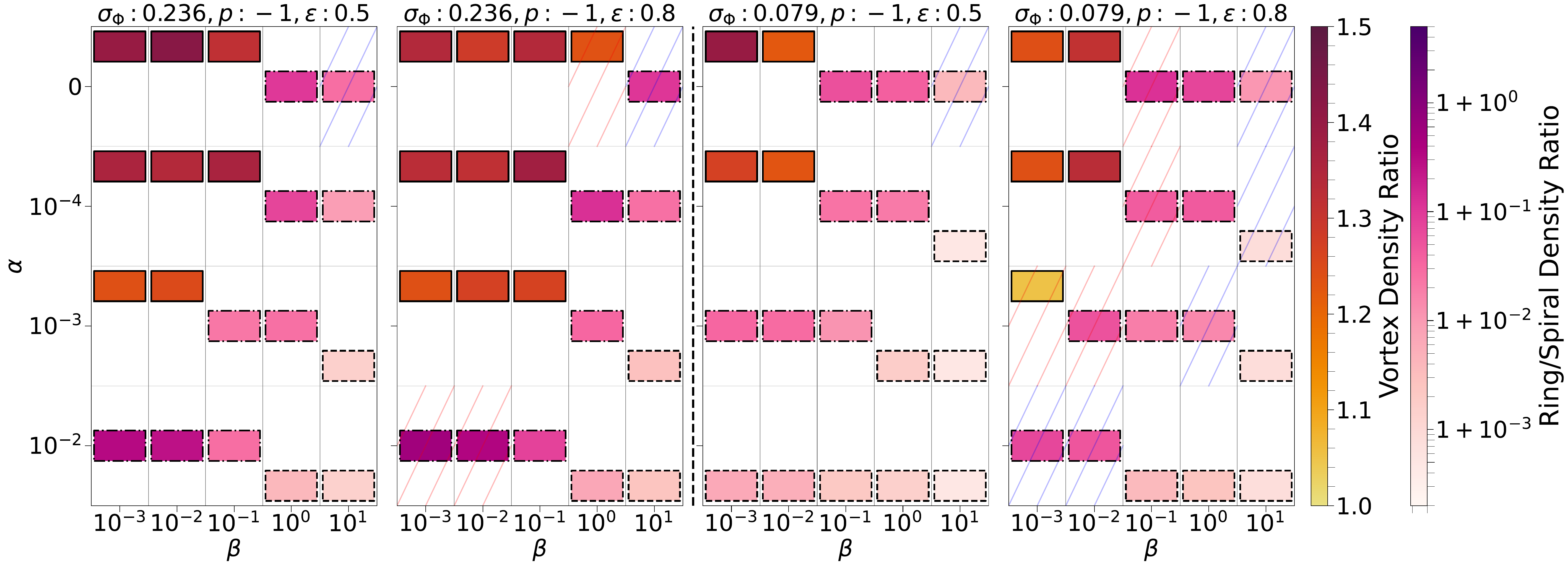}
    \caption{Statistics of shadow-driven substructures based on 80 of 160 simulations with constant disk aspect ratio ($h=0.1$). The figure is divided into two parts by a dashed line, representing the shadow width being 45 degrees on the left and 15 degrees on the right. Within each part, the left column corresponds to disks with $\epsilon=0.5$, while the right column corresponds to disks with $\epsilon=0.8$. Each subfigure in the $\beta$-$\alpha$ sections represents a specific combination of parameters. Within each $\beta$-$\alpha$ section, there are three rows representing the dominant structures in the disk: vortices/crescents, rings, and spirals from top to bottom. Each type of structure is represented by a different type of square marker, whose color represent the density contrast of the substructures. The figure also includes red and blue line shaded areas, indicating disks undergoing transitions from vortex-ring and ring-spiral phases, respectively.}
    \label{Fig:shadow_plot_int_sim}
\end{figure*}

The simplified statistical results are presented in Figure \ref{Fig:shadow_plot_int_sim}, and more detailed ones are provided in Figures \ref{Fig:shadow_plot_int} and \ref{Fig:shadow_plot_azw_and_pa}. It is important to emphasize that panels shaded with red or blue lines are actually undergoing a vortex-ring transition or a ring-spiral transition state (see discussion in Appendix \ref{App:transition state}). 

Generally speaking, shadows are capable of generating different kinds of substructures under different parameter settings. Additionally, we found that the dominant form of shadow-driven substructures changes from spirals to rings and eventually becomes vortices/crescents as cooling timescales and/or viscosity decreases. Where exactly the transition occurs depends on other parameters such as the shadow amplitude, width, and disk aspect ratio, etc., and these will be discussed in more detail in the following subsections.

\subsection{Statistics for Spirals} \label{subSec:sts_spiral}

Two-arm spirals are fundamental substructures in our simulations, dominating in disks with cooling timescales significantly longer than the dynamical timescale, high viscosity ($\alpha > 10^{-3}$), or very weak shadow amplitude (see Section \ref{Sec:discussion}). Here, we focus on discussing their density contrast, pattern speed, and pitch angle.

--{\it Density contrast}. In general, stronger thermal forcing, higher shadow amplitude, wider shadow width, etc. leads to stronger density contrast in the spirals. However, as the spiral-dominated regime generally requires weak thermal forcing, the spiral amplitudes are typically low (with upper limit only $1\%$ higher than background density at the same radius).

--{\it Pattern speed}. Spirals found in our simulations are density wave patterns with zero pattern speed, which also results in non-precessing rings. More generally, the spiral pattern speed exactly matches the shadow's pattern speed, which will be further discussed in Section \ref{subSec:rotating shadow}.

 --{\it Pitch angle}. The pitch angle is solely affected by the disk aspect ratio. With weak thermal forcing, we consider the dispersion relation of spiral density waves in the linear regime under the WKB approximation \citep{Lin1964ApJ...140..646L}
\begin{equation} \label{Equ:dispersion relation}
    m^{2}(\Omega_{p}-\Omega)^{2}=k^{2}c_{s}^{2}+\kappa^{2}.
\end{equation}
Here, $\Omega_{p}$ represents the spiral pattern speed, $k$ is the radial wave number, and $\kappa\approx\Omega_K$ is the epicyclical frequency. The spiral pitch angle can be estimated by $\alpha_{p}=\partial r/(r\partial\phi)\approx m/(|k|r)$. With $\Omega_{p}=0$ and $m=2$, we obtain $\alpha_{p}\sim \frac{2}{\sqrt{3}}h=constant$ for $p=-1$ disks and $\alpha_{p}\sim \frac{2}{\sqrt{3}} h_{0}r^{0.25}$ for $p=-0.5$ disks. Taking the disk parameters used in our simulations (with $h_0=0.1$) and averaging over radius gives $\alpha_{p}=6.6^\circ$ for disks with $p=-1$ and $\alpha_{p}=12.1^\circ$ for disks with $p=-0.5$. These estimated values agree well with our simulation results, which we find to be $7.344_{-0.535}^{+0.607}$$^\circ$ and $13.202_{-1.766}^{+2.16}$$^\circ$ (see Figure \ref{Fig:shadow_plot_azw_and_pa}), respectively.

\begin{figure*}[htbp]
    \centering
    \includegraphics[width=0.8\textwidth]{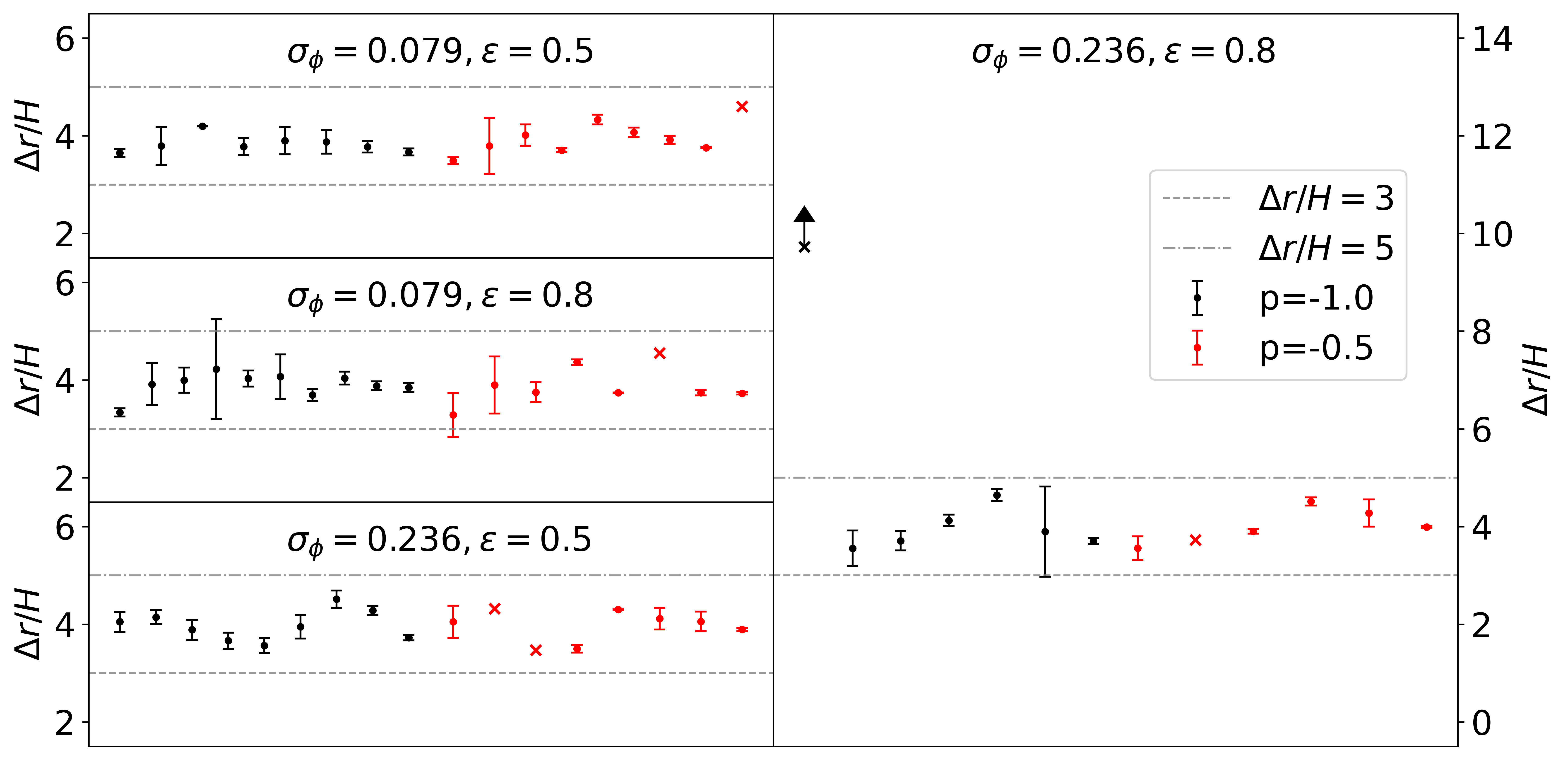}
    \caption{The average and error of normalized ring spacing. The black solid line represents simulations with $p=-1.0$, while the red solid line represents simulations with $p=-0.5$. Points marked with an 'x' do not have error bars. In this case, there are at most two selected rings within the disk. Upward arrow represents lower limit points, where there is only one ring within the disk.}
    \label{Fig:ring_spacing}
\end{figure*}

\begin{figure*}[htbp]
    \centering
    \includegraphics[width=0.8\textwidth]{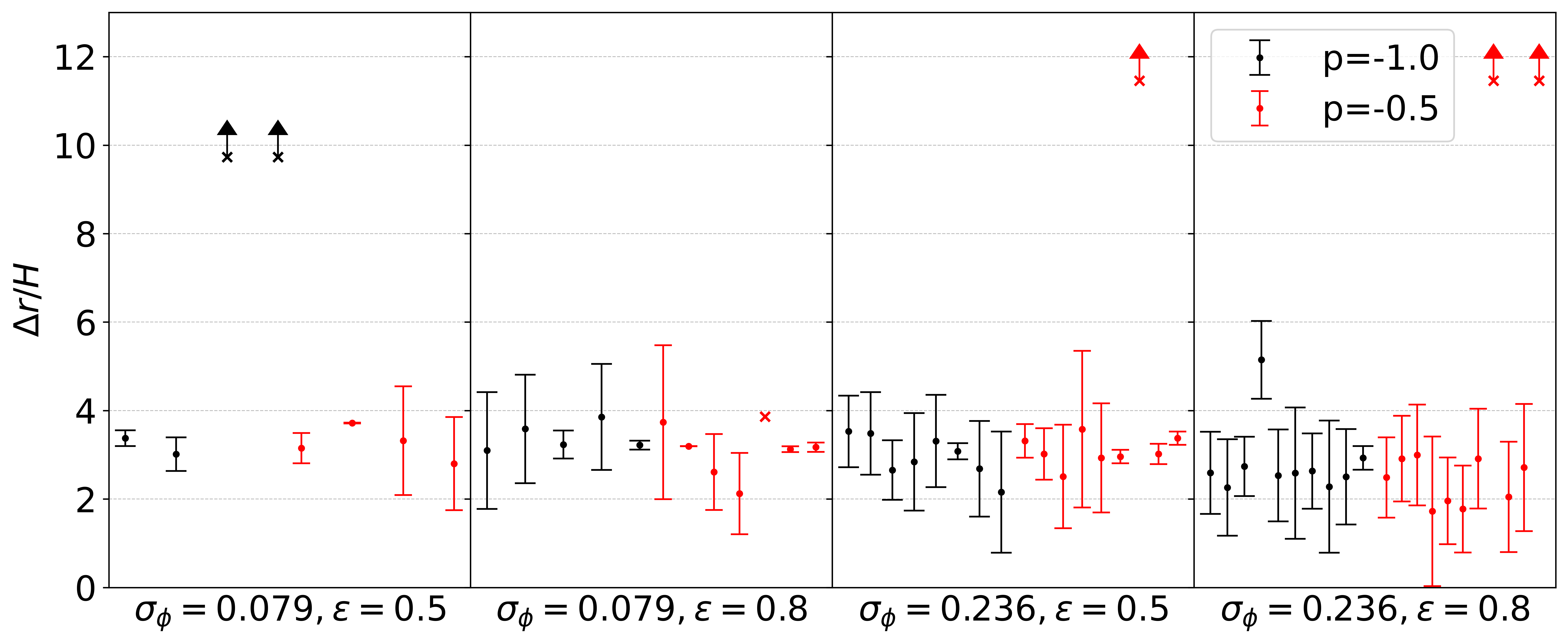}
    \caption{The average and error of normalized vortex/crescent spacing. The black solid line represents simulations with $p=-1.0$, while the red solid line represents simulations with $p=-0.5$. Points marked with an 'x' do not have error bars. In this case, there are at most two selected vortices/crescents within the disk. Upward arrow represents lower limit points, where there is only one selected vortex/crescent within the disk.}
    \label{Fig:vortex_spacing}
\end{figure*}

\subsection{Statistics for Rings} \label{subSec:sts_ring}

In our simulations, rings dominate in disks with cooling timescales comparable to the dynamical timescale ($\beta \sim 1$) when $\alpha$ is roughly below $10^{-3}$. For much higher viscosity, rings dominate even when the cooling rate approaches the isothermal limit ($\beta=10^{-3}$). Typically, this value is $\alpha=10^{-2}$ for disks with $\sigma_{\phi}=0.236$ and $\alpha=10^{-3}$ for disks with $\sigma_{\phi}=0.079$. Overall, the parameter space for the dominance of rings is modest thermal forcing, in between the cases that form vortices/crescents (strong forcing, see next subsection) and spirals (weak forcing). In fact, we pose that rings can be viewed either as ``reconnected spirals" (stated in Section \ref{subSec:ring}), or ``failed vortices", where the latter connection arises from the finding that vortex-ring transitions often involve crescents with very large aspect ratios, although the boundary between this transition is not necessarily clear-cut, and will be further discussed in Appendix \ref{App:transition state}.
Below, we will focus on ``normal" rings (not under transition), and will discuss the density contrast, ring radial width, ring spacing, eccentricity, and the parameters that have strong influence on them.

--{\it Density contrast}. As shown in Figure \ref{Fig:shadow_plot_int_sim} and \ref{Fig:shadow_plot_int}, gas densities are typically $1-20\%$ higher than the background density in ring-dominant disks, and ring density contrast is enhanced by larger shadow amplitude and width. Density contrast could reach very small values, such as $0.3\%$, in the ring-spiral transition, and very large values, such as $50\%$, in the vortex-ring transition.

--{\it Width and spacing}. The ring widths in our simulations are usually 2 times the local scale height, regardless of shadow parameters.
Similarly, for almost all cases, the spacing between neighboring rings is approximately $4H$, as depicted in Figure \ref{Fig:ring_spacing}. There is very small deviations from the mean, indicating a highly uniform distribution of rings within the disk.

--{\it Eccentricity}. As will be stated in Section \ref{Sec:discussion}, ring structures are generated following the ``reconnection" of two-armed spirals in the early stages of disk evolution, causing the ring to become eccentric with zero pattern speed (as shadows are stationary). More flared disk morphology results in larger spiral pitch angles, making the spirals less tightly wound. As a result, the rings formed in this case tend to be more eccentric. Additionally, we find that viscosity has a strong impact on eccentricity. Typically, ring eccentricity varies from 0.1 to 0.7 as $\alpha$ increases from $0$ to $10^{-2}$ in our simulations (see Figure \ref{Fig:shadow_plot_azw_and_pa} for details). The angle between the ring's major axis and the effective shadow center (e.g. $\phi=0^\circ, 180^\circ$ when $\beta=0.001$) is typically between $80^\circ$ and $110^\circ$.

\subsection{Statistics for Vortices and Crescents} \label{subSec:sts_vortex}

As we mentioned in Section \ref{subSec:vortex} and better seen in Figure \ref{Fig:shadow_plot_int_sim}, vortices/crescents tend to dominate in disks characterized by fast cooling processes ($\beta<1$), low viscosity ($\alpha=0,10^{-4}$), high shadow amplitudes ($\epsilon=0.8$), and wide shadow widths ($\sigma_{\phi}=0.236$). Such parameter settings all point to strong thermal forcing. Below, we discuss the properties of the shadow-driven vortices/crescents, focusing on density contrast, spacing and aspect ratio of vortices/crescents under the influence of these parameters.

--{\it Vorticity and density contrast}. The density contrast of substructures is a crucial factor as it directly influences their detectability. From our explorations, the density of the crescents are typically $10-50\%$ higher than the average density at same radius for all vortex-dominated disks. The density contrast is generally slightly higher for stronger shadow intensity, larger shadow width, and faster cooling, but the trend is not definitive given the chaotic nature of the system. 
The normalized vorticity ranges from 0.1 to 0.6 in vortex-dominated disks, with vorticity around 0.2 in most cases, potentially reaching up to 0.6 in the most extreme cases (large $\epsilon$ and $\sigma_{\phi}$). No clear relationship is found between vorticity and density contrast due to the high turbulence level, which is around $0.1c_{s}$. 
The velocity deviations from local Keplerian inside the vortex region ranges from 0.4 to 1.2 $c_{s}$, indicative of strong rotation in the vortices.

--{\it Spacing}. The statistical results of the spacing of vortices/crescents are plotted in Figure \ref{Fig:vortex_spacing}. In all simulations, the distance between neighboring vortices/crescents is typically between $2H$ and $4H$. The spacing is less uniform compared to rings, and is related to the fact that vortex-dominated disks are usually turbulent. Note that the small error bars in a few cases are related to very limited number of vortices/crescents (2 or 3); the lower limit point represents the case where there is only one vortex-induced crescent in the disk. Similar to the case shown in Figure \ref{Fig:vortex}, the azimuthal locations of the vortices/crescents are largely random with no direct correlation with the position of the shadows.

--{\it Aspect ratio}. The aspect ratio of crescents/vortices is less affected by different parameters. Typically, in vortex-dominated disks, this value is about 6. However, for cases close to (for example, $\sigma_{\phi}=0.079$, $p=-1$, $\epsilon=0.5$, $\alpha=0$, $\beta=0.01$) or undergoing (for example, $\sigma_{\phi}=0.236$, $p=-0.5$, $\epsilon=0.8$, $\alpha=10^{-4}$, $\beta=1$) the vortex-ring transition in parameter space, the aspect ratio can be very large (greater than 12). More detailed results are shown in Figure \ref{Fig:shadow_plot_azw_and_pa} in Appendix \ref{App:detail_sts}.

\section{Discussion} \label{Sec:discussion}

\begin{figure*}[htbp]
    \centering
    \includegraphics[width=0.8\textwidth]{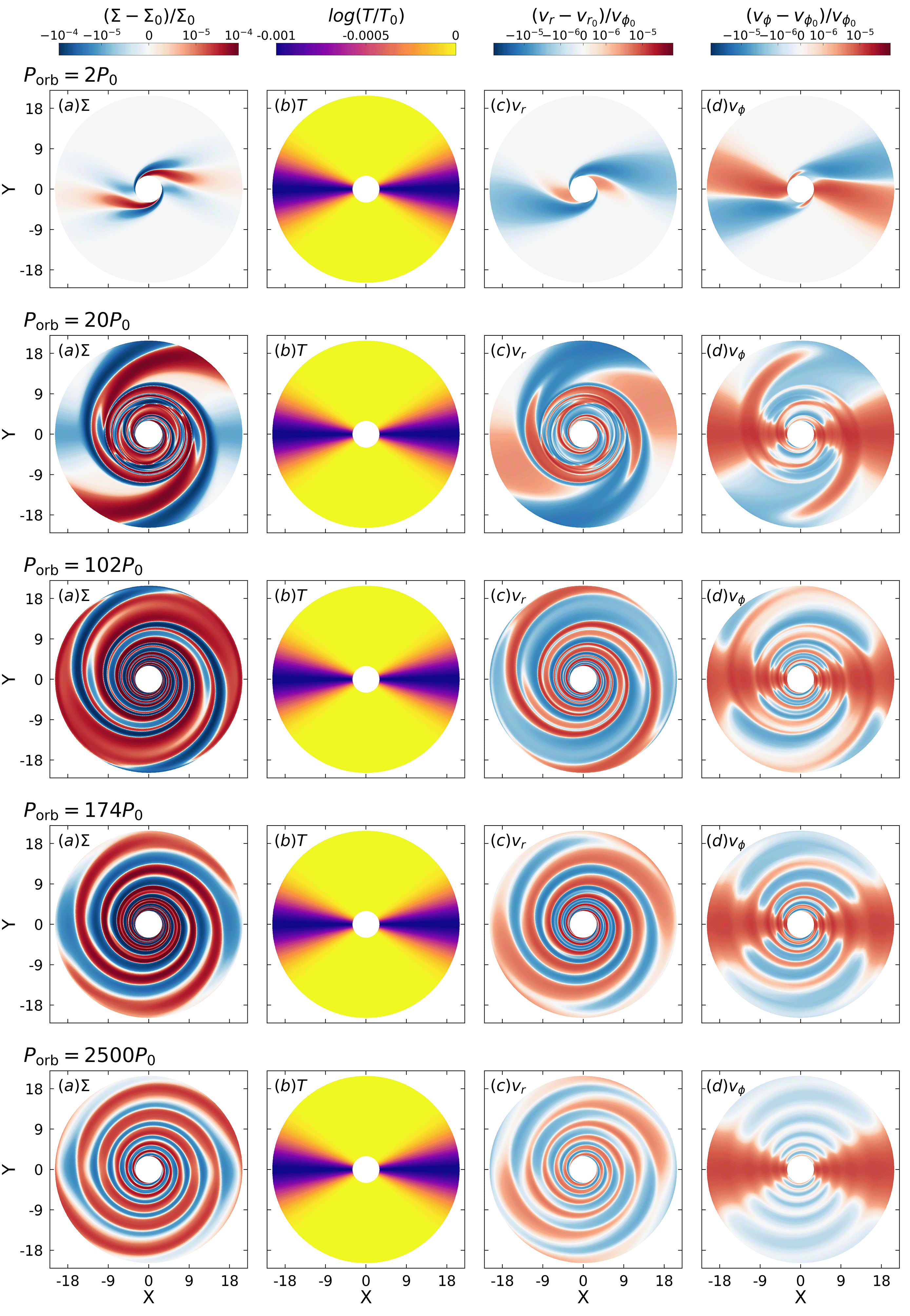}
    \caption{Density (a), temperature (b), radial velocity (c) and azimuthal velocity (d) evolution in linear evolution disk ($\sigma_{\phi}=0.236, \epsilon=0.001, \alpha=0, \beta=0.001, p=-1.0$). All quantities except for $v_{r}$ are normalized by their initial values, and $v_{r}$ is normalized by initial $v_{\phi}$. }
    \label{Fig:linear}
\end{figure*}

\begin{figure*}[htbp]
    \centering
    \includegraphics[width=0.85\textwidth]{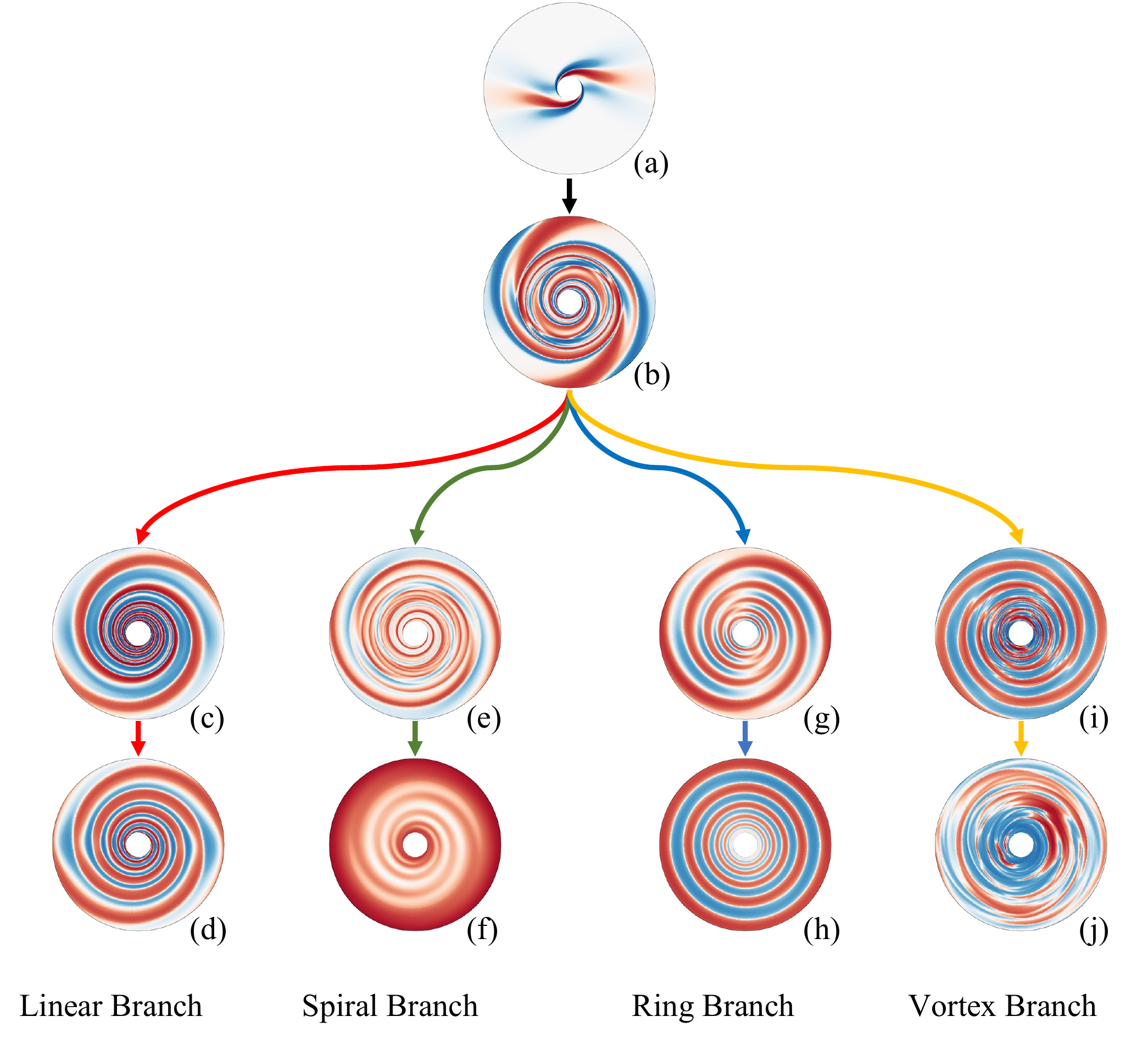}
    \caption{Formation processes of substructures from linear to non-linear regime. All plots show density contrast in disks in logarithmic scale.}
    \label{Fig:formation}
\end{figure*}

In this paper, we have conducted simple numerical experiments to study the dynamical consequence of shadows cast from the inner disk to the outer disk as a result of thermal forcing. We have restricted ourselves to a small number of parameters, and the discussion has been largely phenomenological. In this section, while not going into full detail, we conduct additional studies to help better understand the origin and trend of shadow-driven substructures, and briefly discuss their potential implications.

\subsection[]{Linear regime}\label{subSec:linear regime}

Based on the analysis and discussions in the previous sections, here we provide further analysis to gain better physical insights on the shadow-driven substructure formation. As we observe that in all cases, substructure formation starts from the formation of two-armed spirals under our shadow prescriptions. This suggests that spirals are the most fundamental form of shadow-driven substructure, and it can be instructive to look into how spirals form and evolve under very weak thermal forcing to avoid nonlinear effects. We thus further conducted a series of 2D inviscid hydrodynamic simulations with varying perturbation strengths ($\epsilon=0.001$, $\epsilon=0.01$, $\epsilon=0.1$) while keeping the cooling timescales consistent ($\beta=0.001$). Without viscosity, the simulations are in hydrostatic equilibrium to start with before thermal forcing is introduced. 

In Figure \ref{Fig:linear}, we present the results from the $\epsilon=0.001$ simulation (run L-hm-S-NR in Table \ref{Tab:table-1}).
When the shadow is introduced, gas flows into the shadowed region in a counterclockwise manner. The gas between the shadow center (pressure minimum) and its rear edge, i.e., between $315^\circ$ and $360^\circ$ in the first row of Figure \ref{Fig:linear}, gets accelerated, while the gas between the shadow center and its leading edge, i.e., between $0^\circ$ and $45^\circ$ in the first row of Figure \ref{Fig:linear}, gets decelerated. This leads to gas piling up near the shadow center, while the neighboring gas is slightly rarefied, which naturally launch density waves.

As the disk evolves, such density waves wind up due to differential rotation (see second row of Figure \ref{Fig:linear}). 
In the meantime, the periodic forcing at the shadow location continues, keep launching new density waves, leading to interference. After a few local orbits, the system reaches a relatively steady pattern of two-arm spirals (see third and forth rows of Figure \ref{Fig:linear}), which remain stable over long-term. The spirals share the same pattern speed of the shadows (in this case, zero), and the pitch angle also remains unchanged. We note that this is very different from planet-induced spirals in that a planet launches density waves through discrete Lindblad resonances, while as shadows are cast over a wide range of radii, each radius can excite its own density waves. In our case, the pattern speed of the shadow is zero, and the only relevant resonance condition is simply given by $\Omega=\kappa/m$, where $m=1, 2, \cdots$. However, taking $m=2$, we see that with $\kappa\approx\Omega$ for Keplerian disks, no resonance condition is satisfied. In other words, the two-armed spirals are not driven by Lindblad resonances, but are the effective eigen-state of thermally-forced oscillations.

\subsection[]{Towards the nonlinear regime}\label{subSec:nonlinear regime}

We note that even in the linear regime, the spiral patterns are distorted due to thermal forcing. These can be most easily seen from the velocity perturbations in the last three columns of Figure \ref{Fig:linear}. They are also present in the density perturbations where the amplitude of the spirals varies across the shadow region. The form of the distortion can depend on system parameters, which is found to be different in Figure \ref{Fig:spiral} where cooling time is significantly longer. We speculate that such distortions are the source of instability when thermal forcing enters the nonlinear regime.

Based on our discussions in the previous sections, we summarize the formation of shadow-driven substructures in Figure \ref{Fig:formation}. Irrespective of whether thermal forcing is linear or nonlinear, the initial phase of the development is similar, involving the formation of two-armed spirals, as shown in (a)-(b). The spirals persist under linear and weakly nonlinear thermal forcing, as seen in the ``linear branch" and ``spiral branch" in (c)-(f). The properties of spirals are similar between the linear and weakly nonlinear regimes, in terms of pitch angle and pattern speed.

When thermal forcing becomes slightly stronger, the spiral arms undergo a relatively quiescent transformation by ``reconnecting" into eccentric rings (see Figure \ref{Fig:formation}(g), (h)). The eccentricity of these rings is largely set by the pitch angles of the original two-arm spirals stage and disk viscosity.  However, when the thermal forcing becomes too strong, the spirals break in a highly chaotic manner (see Figure \ref{Fig:formation}(i), (j)), leading to the formation of more localized vortices/crescents. 

\begin{figure*}[htbp]
    \centering
    \subfigure[]{
    \label{subFig:Omega1}
    \includegraphics[width=0.3083\textwidth]{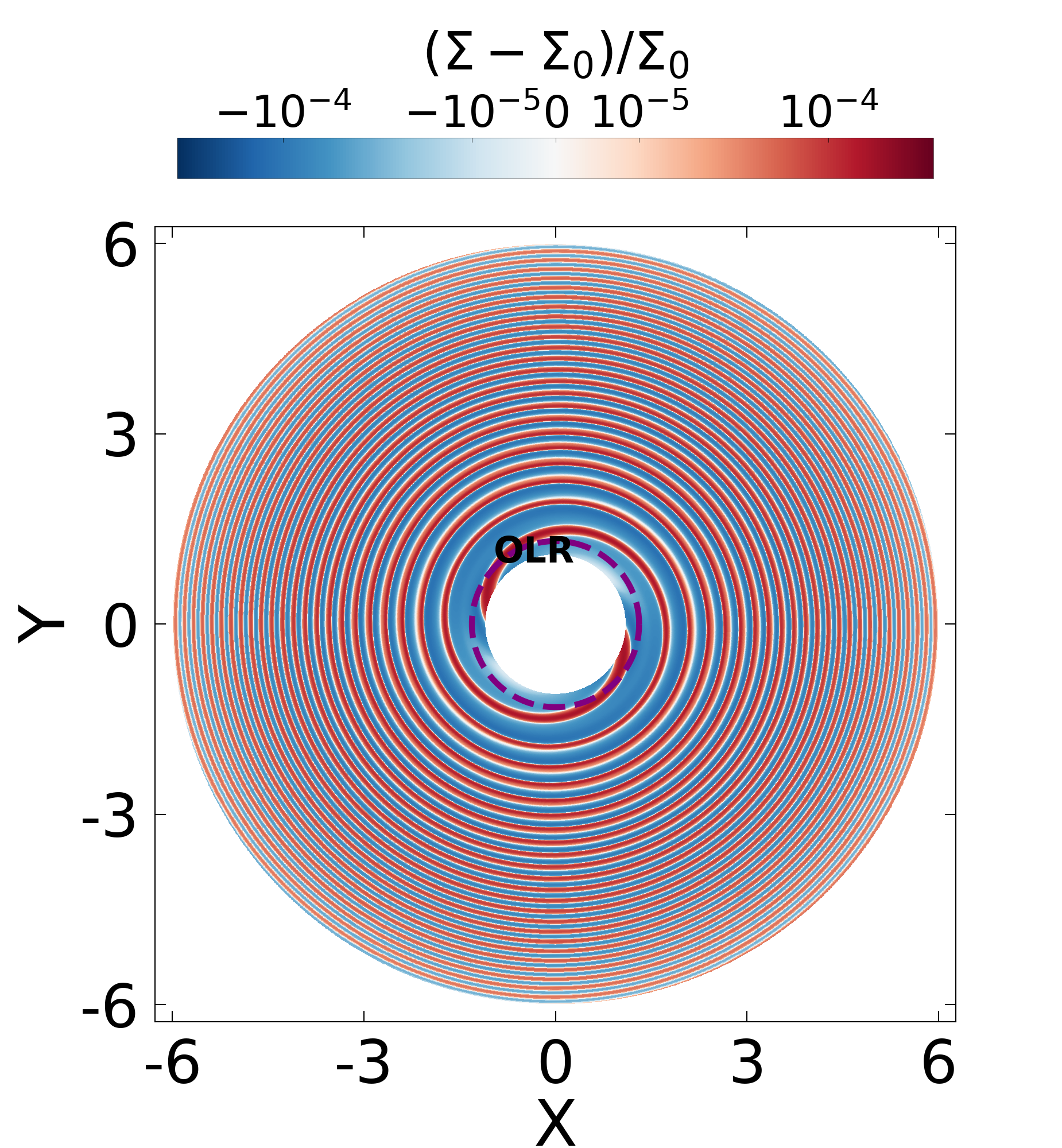}}
    \hspace{0.1cm}
    \subfigure[]{
    \label{subFig:Omega0.03}
    \includegraphics[width=0.3\textwidth]{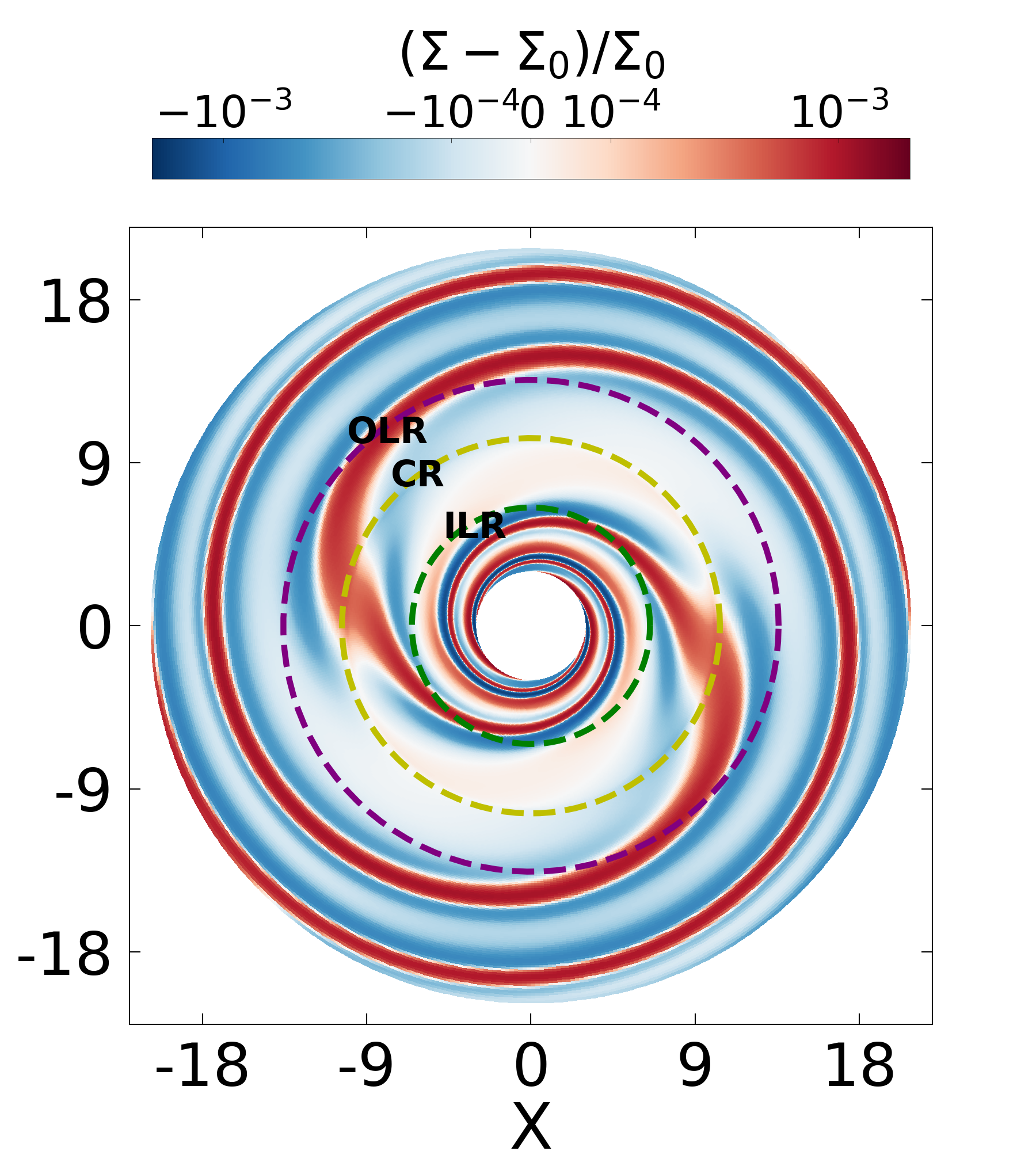}}
    \hspace{0.1cm}
    \subfigure[]{
    \label{subFig:Omega0.003}
    \includegraphics[width=0.3\textwidth]{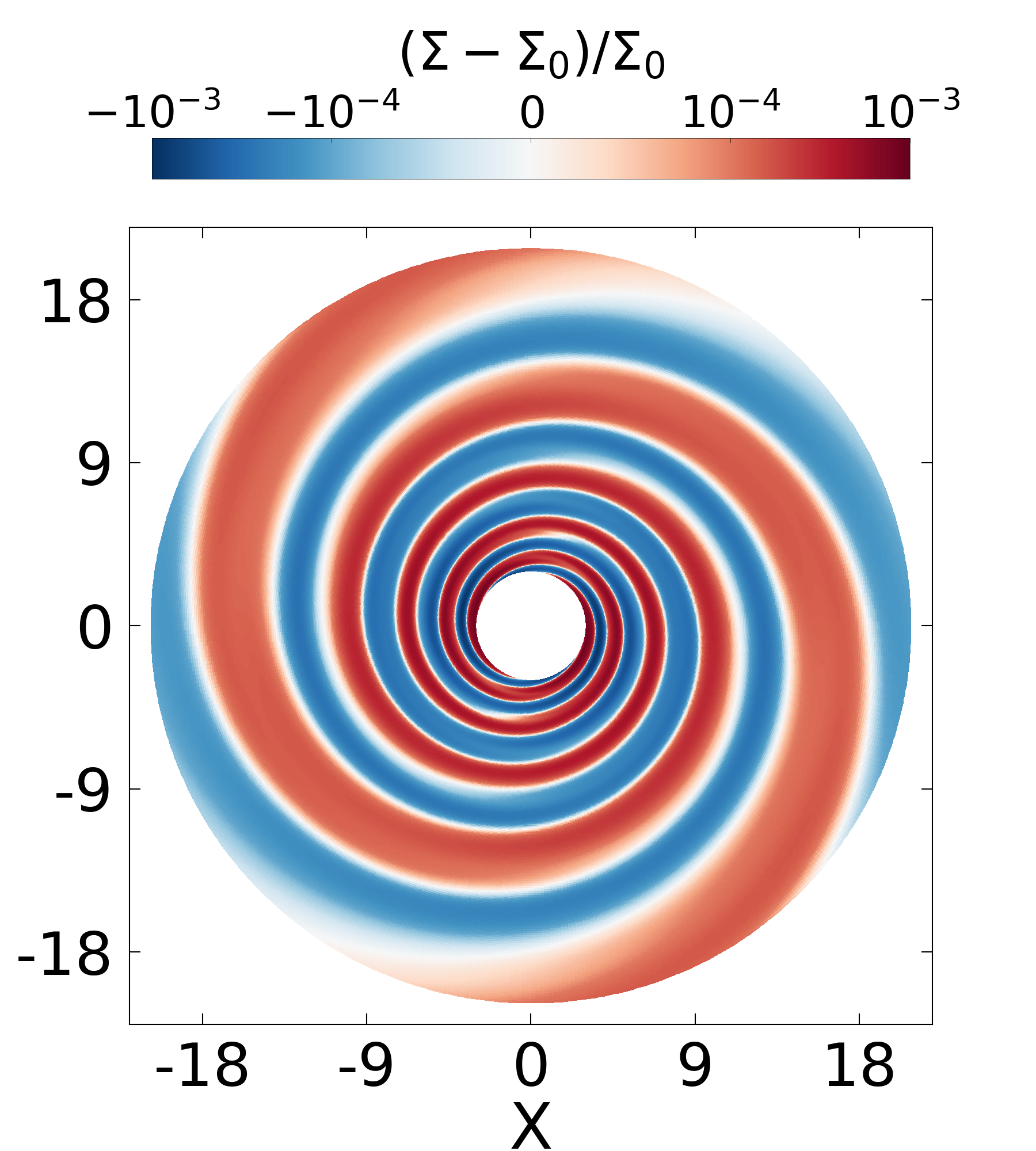}}
    \caption{Normalized density perturbation for simulations with $\Omega_{\text{shadow}}=1\Omega_{0}$ (\ref{subFig:Omega1}), $0.03\Omega_{0}$ (\ref{subFig:Omega0.03}), and $0.003\Omega_{0}$ (\ref{subFig:Omega0.003}) at the final steady state under the parameters setting as $\sigma_{\phi}=0.236, \epsilon=0.001, \alpha=0, \beta=0.001, p=-1.0$. The radii of corotation resonances (CR), inner Lindblad resonances (ILR), and outer Lindblad resonances (OLR) are shown as yellow, green, and purple dashed lines respectively.}
    \label{Fig:rotating shadow}
\end{figure*}

\subsection[]{Rotating shadows}\label{subSec:rotating shadow}

In this paper, we have only discussed the situation when the shadow's pattern speed is zero. However, if the misaligned inner disk precesses around the central star, the shadow cast from the inner region would have a pattern speed, which then changes the resonance condition discussed in Secton \ref{subSec:linear regime}. To extend our study to more general conditions, we have conducted additional simulations with rotating shadows in the linear regime, with three different shadow pattern speeds, $\Omega_{\text{shadow}}=1\Omega_{0}$ (run L-hm-S-FR), $0.03\Omega_{0}$ (run L-hm-S-MR), and $0.003\Omega_{0}$ (run L-hm-S-SR). Here, $\Omega_{0}$ is Keplerain angular velocity at $r=1$. The detailed parameter settings can be found in Table \ref{Tab:table-1}. The density structure from these simulations in the final states are shown in Figures \ref{Fig:rotating shadow}. We measure the pattern speed of the spirals $\Omega_{p}$ in these situations, and we confirm that in all three cases, the spirals all have $\Omega_{p}=\Omega_{\text{shadow}}$.

Given the pattern speed, the radii of corotation resonances (CR), inner Lindblad resonances (ILR), and outer Lindblad resonances (OLR) can be calculated by $\Omega_{p}=\Omega$, $\Omega_{p}=\Omega\pm\kappa/m$ (with $m=2$), and shown as yellow, green, and purple dashed lines in Figures \ref{Fig:rotating shadow}. With the WKB dispersion relation \ref{Equ:dispersion relation}, the permitted regions for density wave propagation are outside the Lindblad resonances.
In the fast-rotating case $\Omega_{p}=\Omega_0$, density waves are permitted beyond the OLR, and the spirals are tightly wound towards outer radii with pitch angle $\alpha_{p}\sim h(\Omega/\Omega_p)\sim h(r/r_0)^{-3/2}$. Even with a resolution of $N=2048$ in Figure \ref{subFig:Omega1}, it is still insufficient to resolve the spirals across the entire disk, weakening the spirals at the outer disk by numerical dissipation. With intermediate $\Omega_{\rm shadow}=0.03\Omega_0$, the ILR and OLR are located at $r=6.5$ and $r=13.5$, respectively. Clearly, there are well-defined spirals outside the Lindblad resonances, which break inside the Lindblad resonances. In the slow-rotating case with $\Omega_{\rm shadow}=0.003\Omega_0$, even the ILR is beyond the computational domain, and the results are largely identical to the stationary case described in Section \ref{subSec:linear regime}.

Given the discussion above, we expect the results presented in this paper largely applies to regions inside the ILR in slowly-precessing shadows. Although not the focus of this paper, it is worth noting the significance of moderately rotating shadows, where the corotation radius lies within the disk region. Our findings are morphologically similar to those of \cite{Montesinos2018MNRAS.475L..35M}, who demonstrated that the morphology of shadow-driven spirals notably resembles the planetary wakes caused by embedded planets in the disc using radiative transfer. For better comparison with planet-induced spirals, more detailed investigation with more realistic physics (especially dust and radiative processes) is necessary for the slow-rotating case, especially in regions between the ILR and OLR.

\subsection[]{Dependence on disk aspect ratio}
\label{subSec:different aspect ratio}

\begin{figure*}[htbp]
    \centering
    \subfigure[]{
    \label{subFig:h0.05_S}
    \includegraphics[width=0.325\textwidth]{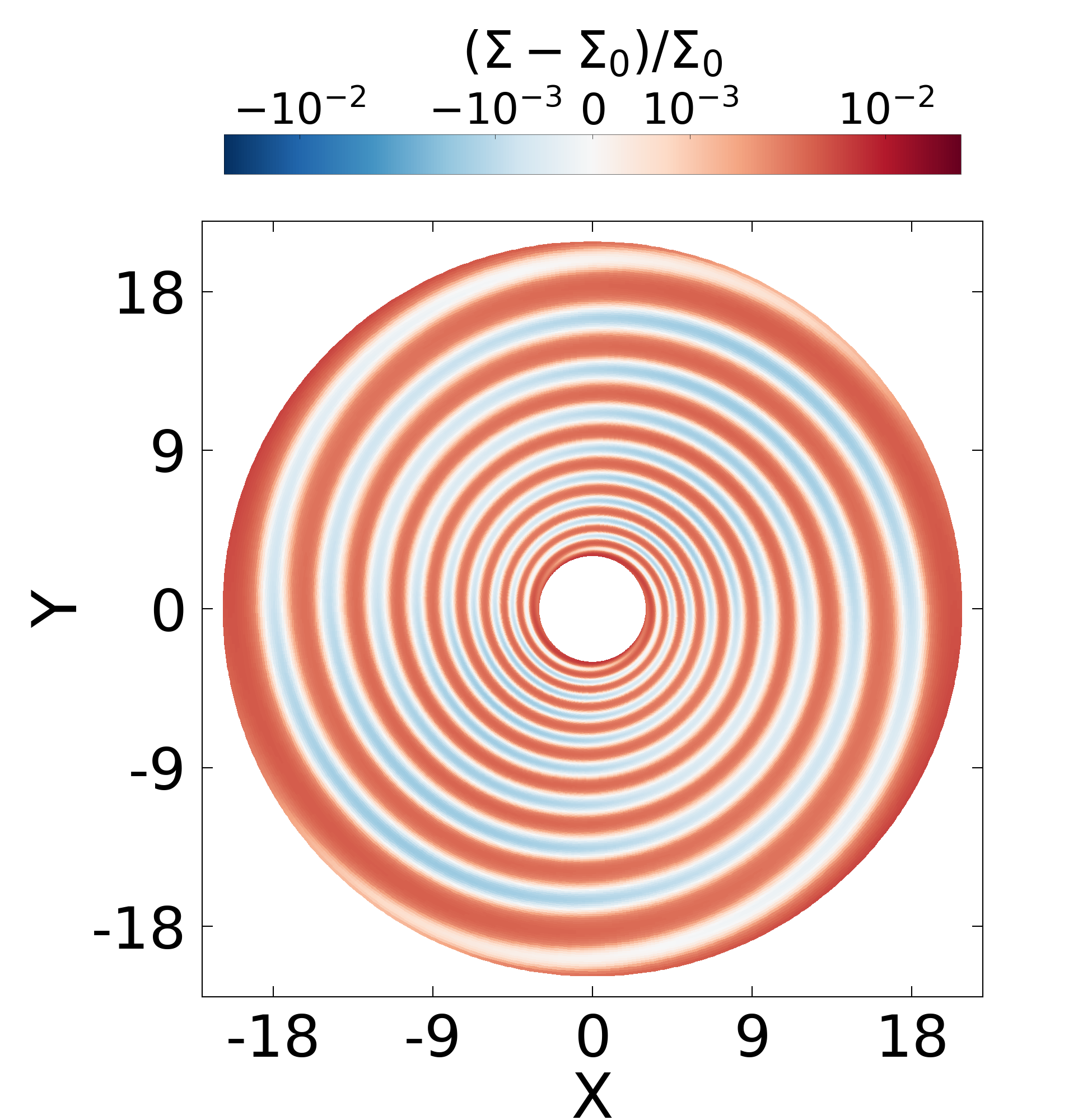}}
    \hspace{0.1cm}
    \subfigure[]{
    \label{subFig:h0.05_R}
    \includegraphics[width=0.3\textwidth]{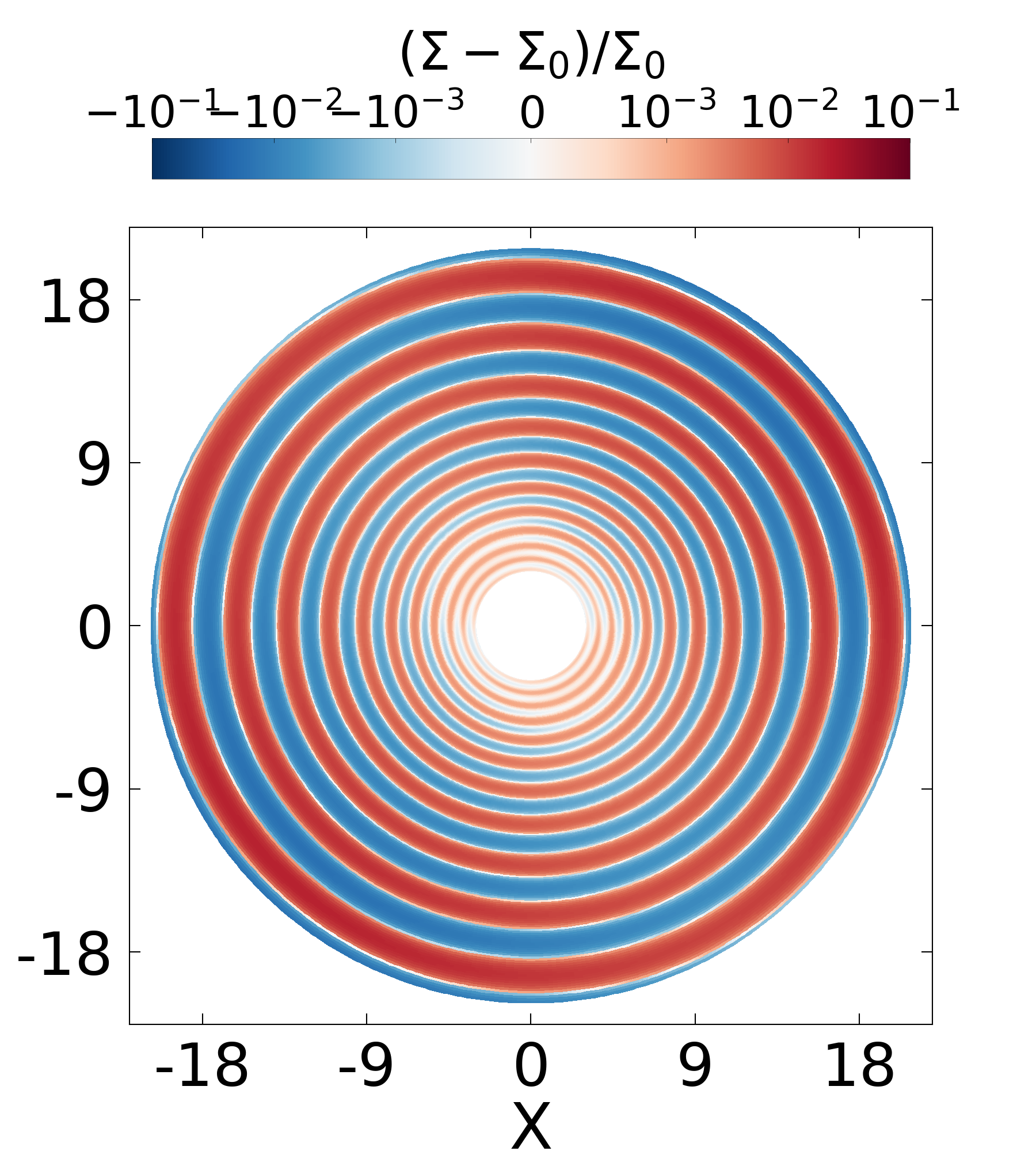}}
    \hspace{0.1cm}
    \subfigure[]{
    \label{subFig:h0.05_V}
    \includegraphics[width=0.3\textwidth]{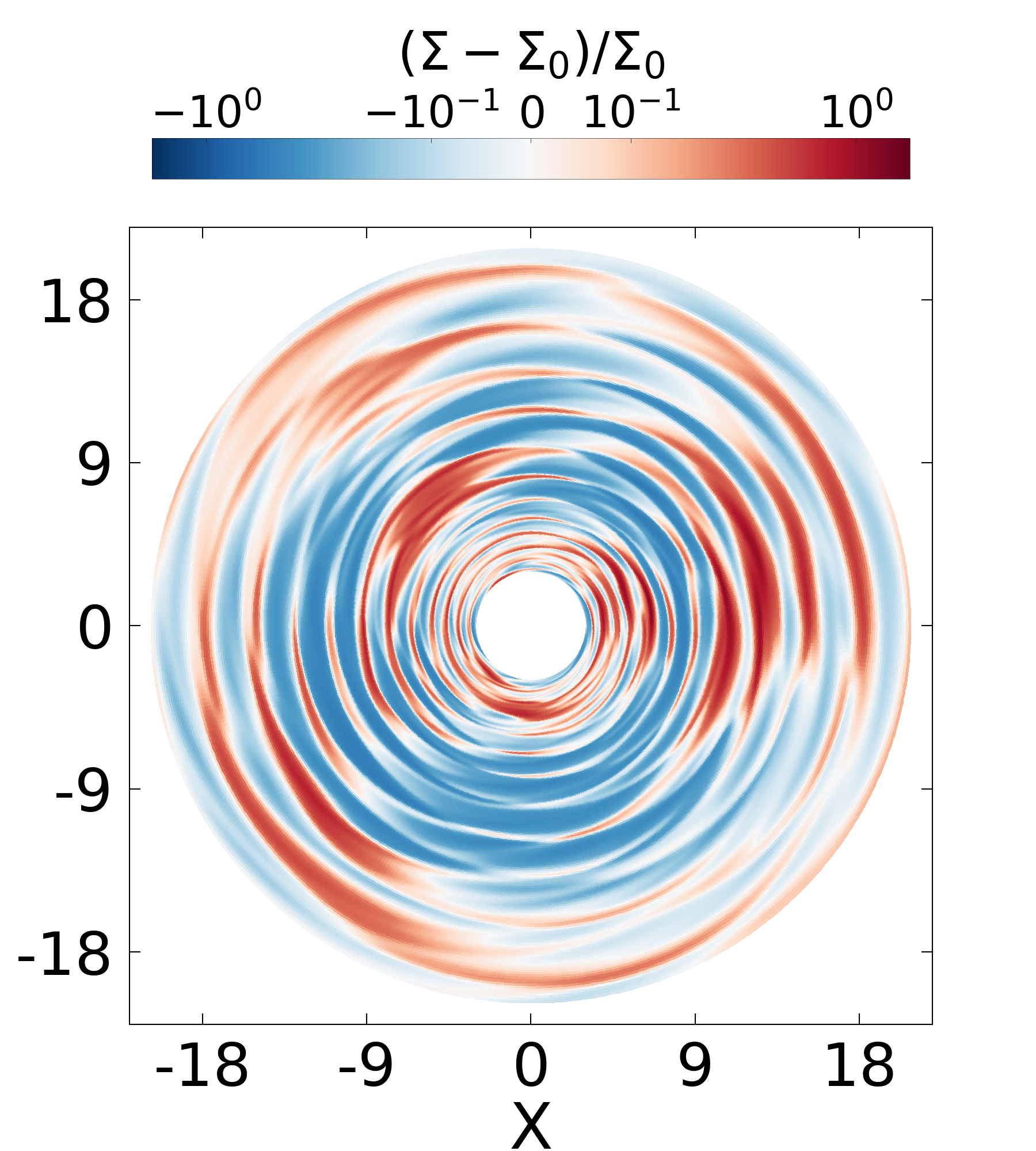}}
    \\
    \subfigure[]{
    \label{subFig:h0.2_S}
    \includegraphics[width=0.325\textwidth]{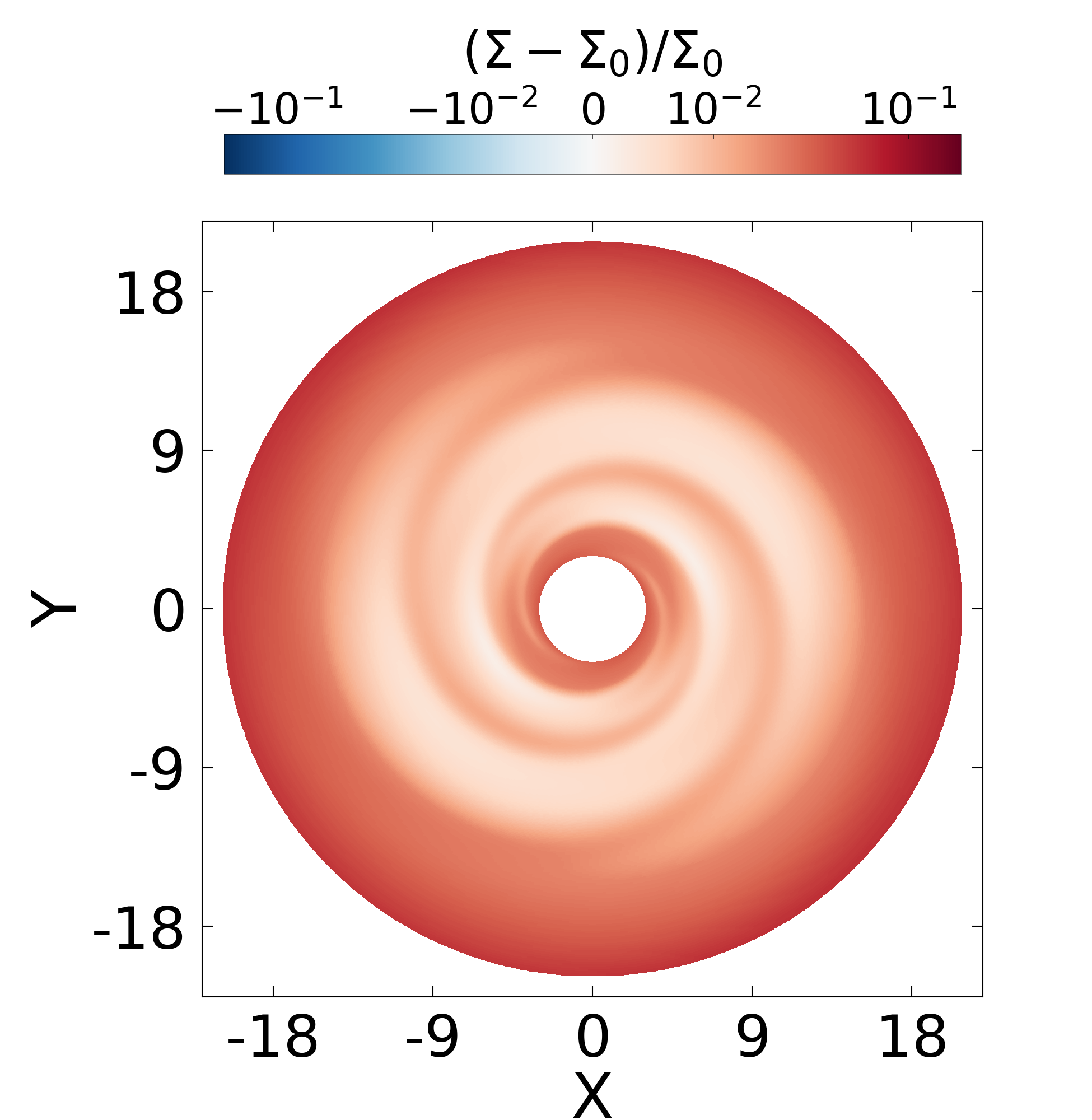}}
    \hspace{0.1cm}
    \subfigure[]{
    \label{subFig:h0.2_R}
    \includegraphics[width=0.3\textwidth]{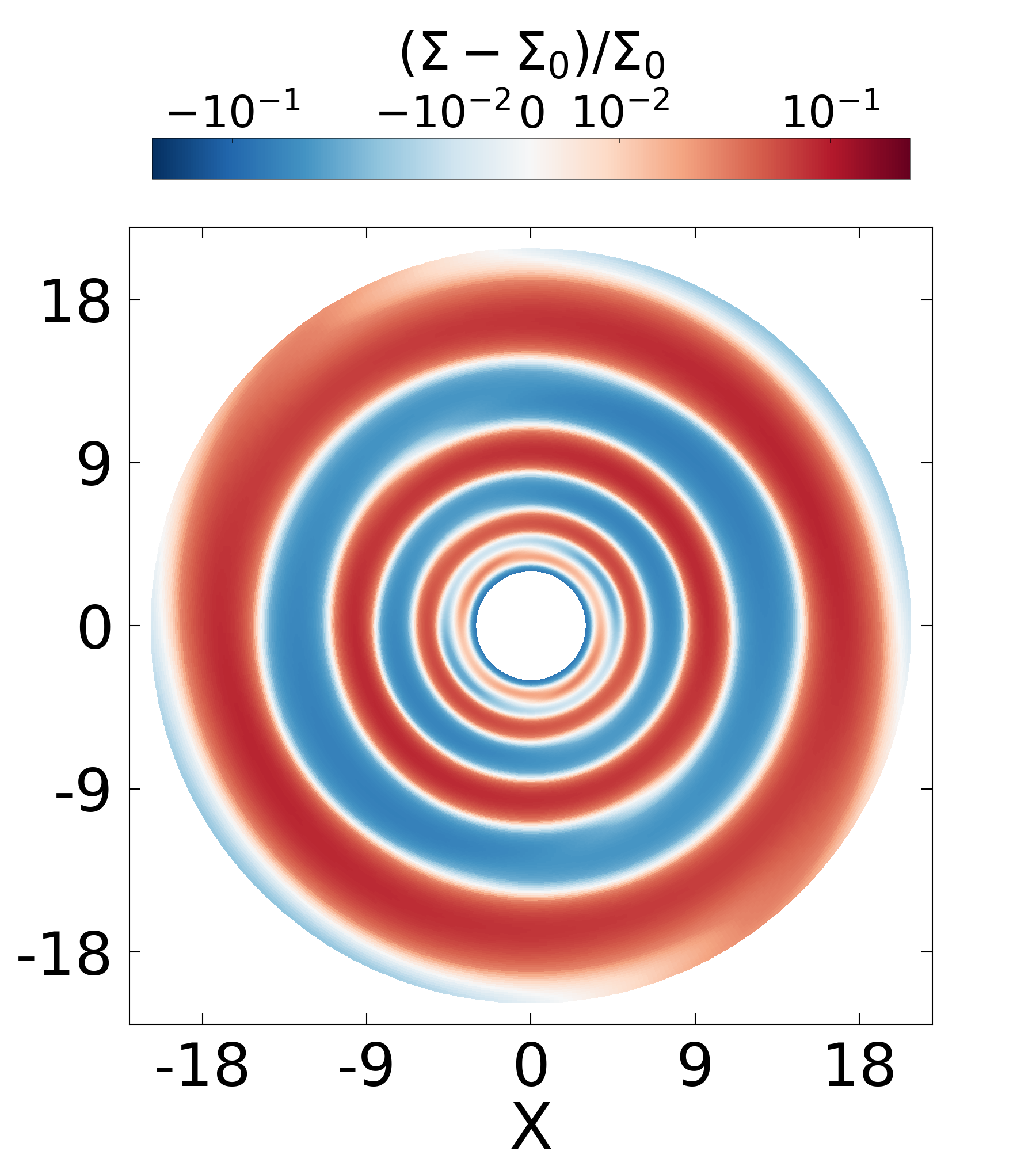}}
    \hspace{0.1cm}
    \subfigure[]{
    \label{subFig:h0.2_V}
    \includegraphics[width=0.3\textwidth]{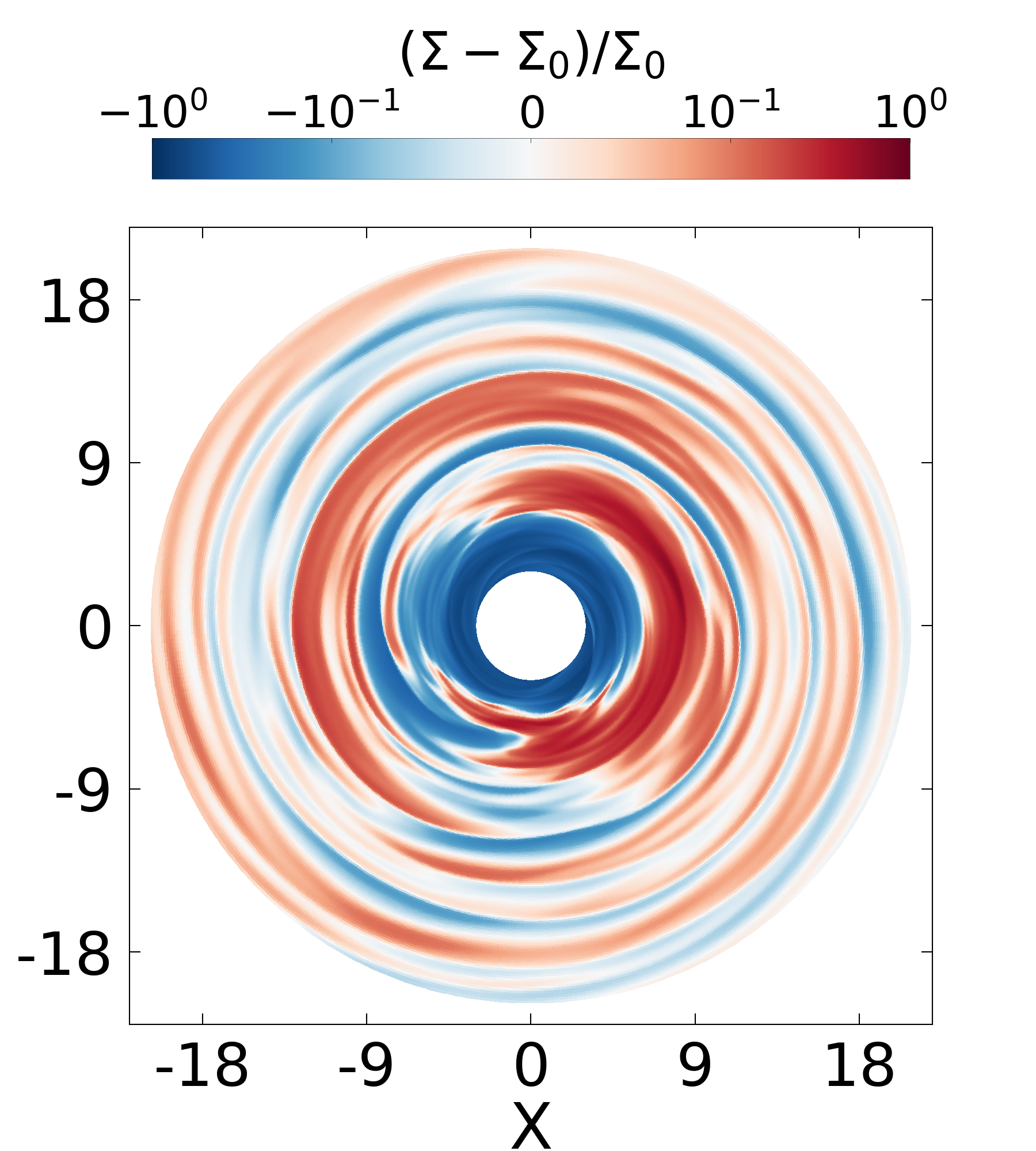}}
    \caption{
    % Normalized density perturbation for runs with different disk aspect ratio.
    Normalized density perturbation for runs with different disk aspect ratios. The parameters are identical to those in the representative runs (NL-hm runs) in Section \ref{Sec:Results}, except for the disk aspect ratio. Detailed parameter settings are provided in Table \ref{Tab:table-1}.
    (a): snapshot with $P_{\rm orb}=20000P_{0}$ for run NL-hs-S-NR ($h_{0}=0.05, \sigma_{\phi}=0.236, \epsilon=0.8, \alpha=10^{-3}, \beta=10, p=-1.0$). (b): snapshot with $P_{\rm orb}=20000P_{0}$ for run NL-hs-R-NR ($h_{0}=0.05, \sigma_{\phi}=0.236, \epsilon=0.5, \alpha=10^{-4}, \beta=1, p=-1.0$). (c): snapshot with $P_{\rm orb}=20000P_{0}$ for run NL-hs-V-NR ($h_{0}=0.05, \sigma_{\phi}=0.236, \epsilon=0.5, \alpha=0, \beta=0.001, p=-1.0$). (d): snapshot with $P_{\rm orb}=3000P_{0}$ for run NL-hl-S-NR ($h_{0}=0.15, \sigma_{\phi}=0.236, \epsilon=0.8, \alpha=10^{-3}, \beta=10, p=-1.0$). (e): snapshot with $P_{\rm orb}=10000P_{0}$ for run NL-hl-R-NR ($h_{0}=0.15, \sigma_{\phi}=0.236, \epsilon=0.5, \alpha=10^{-4}, \beta=1, p=-1.0$). (f): snapshot with $P_{\rm orb}=1500P_{0}$ for run NL-hl-V-NR ($h_{0}=0.15, \sigma_{\phi}=0.236, \epsilon=0.5, \alpha=0, \beta=0.001, p=-1.0$).}
    \label{Fig:diff aspect ratio}
\end{figure*}

In the preceding discussion, we observed that shadow-driven substructures are closely tied to thermal forcing, which is influenced not only by the cooling process but also by disk temperature. Additionally, detailed characteristics of substructures, such as pitch angle or eccentricity, are affected by the disk aspect ratio h. Therefore, it is natural to further investigate the influence of $h_{0}$. We conducted additional simulations with $h_{0}$ ranging from 0.03 to 0.15, focusing on $h_{0}=0.05$ and $h_{0}=0.15$. These simulations, denoted as NL-hs-S-NR, NL-hs-R-NR, NL-hs-V-NR and NL-hl-S-NR, NL-hl-R-NR, NL-hl-V-NR respectively, maintained the same parameters as the representative runs discussed in Section \ref{Sec:Results} except for $h_{0}$ (see Table \ref{Tab:table-1}). We note that here ``S", ``R", and ``V" do not necessarily indicate dominant form of substructures but rather serve to guide the reader that these runs only vary $h_{0}$ compared to representative runs.

In the NL-hs run series ($h_{0}=0.05$), with lower target temperature, we see that the NL-hs-S-NR (Figure \ref{subFig:h0.05_S}) and NL-hs-R-NR (Figure \ref{subFig:h0.05_R}) runs maintain spirals and rings as the primary substructure, respectively. We see the spirals are more tightly wound and the rings spacing remains uniform except for being smaller. The changes are exactly in proportion to $h_0$, and the general properties of the rings and spirals are otherwise identical to those discussed in the NL-hm runs. For the NL-hs-V-NR run, while the vortices are clearly the dominant, many of the overdensities close a full circle, and we identify this run as in the vortex-ring transition state.

In the NL-hl run series ($h_{0}=0.15$), with higher target temperature, we see that all three NL-hl runs retain their spiral, ring and crescent/vortex as the dominant substructure, respectively. Similarly, the spirals are more open, the rings are more eccentric, and the vortices are larger and more widely spaced, as expected. 

Overall, we find that varying $h$ slightly alters the boundary where different forms of substructures dominate, while the general properties for individual substructures largely remain consistent with what we have found in the fiducial simulations with $h_{0}=0.1$.

\subsection[]{Observational implications}
\label{subSec:Observation}

Given the diverse dynamical consequence of shadowing, such disks is expected to exhibit a variety signatures that are potentially observable. However, it should be noted that our work serves as a general exploration without detailed modeling, including radiation transport, dust dynamics, shadow precession rates \citep[e.g.,][]{Pinilla2015,Stolker2016A&A...595A.113S,Wolff2016ApJ...818L..15W,Debes2017}, and realistic shadow morphologies may differ from our prescription \citep[e.g.,][]{Muro-Arena2020,Debes2017}. Additionally, as there are a variety of other mechanisms that can drive substructures \citep[e.g., see reviews by][]{Andrews2020ARAA,Bae+PPVII,Benisty2022_PPVII}, such as planet-disk interactions, icelines etc. Our shadowed disk simulations implicitly assumed a smooth disk to start with, and it is conceivable that the final outcome is set by the interplay between the existing substructures and shadowing. Besides such dynamical interplay, substructures themselves can self-shadow \citep[e.g.,][]{Zhang2021ApJ...923...70Z}, which can further complicate the situation. Therefore, a systematic observational comparison with specific sources is beyond the scope of this work. Below, we mainly discuss general aspects of potential observational implications.

--{\it Spirals}. Spirals generated from shadows may not be easily detectable in the submm continuum or in kinematics, but may be observable in scatter light. Nearly all spiral-dominant disks correspond to weak thermal forcing, resulting in only about $0.1\%$ higher gas density than the background. This not only makes pressure variations across the spirals small that is difficult for dust trapping, and only sufficiently small particles with a stopping time shorter than the spiral crossing time (typically requiring the Stokes number much less than 0.1) can potentially be trapped by the spiral \cite[e.g.][]{Sturm2020A&A...643A..92S,Speedie2022ApJ...930...40S}. 
With the weak spirals, the gas velocity is found to show very small deviationsfrom Keplerian ($\sim 0.1\%$ $v_{k}$, as opposed to $\gtrsim0.5\%$ $v_{k}$ for typical ALMA observations \citep{Pinte2023ASPC..534..645P}.), making it difficult for kinematic detection. On the other hand, such spirals may be detectable in scattered light, as suggested by \cite{Montesinos2016} for the HD 142527 disk, thanks to azimuthal variation of disk scale heights across the spirals, though three-dimensional simulations are needed for proper characterization.

--{\it Rings}. For full disks, our simulations predict the presence of multiple gas rings that are uniformly spaced and weakly eccentric. The relatively high density contrast in our simulations suggests that these rings likely concentrate dust, making them readily observable in sub-mm wavelengths. 
While the resulting dust rings formed are also likely uniformly spaced, whether they can be eccentric remains uncertain (as the eccentric gas ring is a pattern and does not reflect real motion), requiring simulations incorporating dust dynamics. From all simulations, we find that the azimuthal temperature contrast in the ring-dominant disks are typically greater than $8\%$ and can reach up to $50\%$ as they approach to vortex-ring transition in disks with high viscosity and rapid cooling. Such azimuthal temperature variations should result in azimuthal brightness variations in the mm continuum image, which however has not been revealed in in real shadowed disks with rings (e.g., HD 143006). This suggests that thermal forcing by shadows in these systems are likely not as strong as given in our prescriptions, but we caution that without detailed modeling of shadow morphology, radiation transport and dust dynamics, we cannot make specific predictions for individual systems. On the other hand, we comment that both the weakly eccentric ring pattern and low-level of azimuthal temperature variation, if present, may affect the interpretation of azimuthal asymmetries seen in multi-ring systems \cite[e.g.][]{Doi2021ApJ...912..164D,Liu2022SCPMA..6569511L}.
Finally, we note that detection by kinematic signatures, with velocity disturbances being $\sim1\%$ of the Keplerian velocity, is possible but challenging since they are close to ALMA's detection limits.

--{\it Crescents}. 
Vortices generate significant velocity perturbations and are favored sites for dust trapping. 
Given the adopted turbulent viscosity parameter $\alpha_{t}\gtrsim10^{-2}$ in most vortex-dominated simulations, dust with Stokes number $St > \alpha_{t} \sim 10^{-2}$ is expected to concentrate inside vortices overcoming turbulent diffusion \citep{Birnstiel2013A&A...550L...8B}, and can be readily observable in sub-millimeter wavelengths \citep{Zhu2014ApJ...785..122Z}. Previous studies have found that detecting kinematic signatures of vortices can be possible but challenging \citep{Huang/Ping2018ApJ...867....3H},
despite the relatively large vorticity (typically around $0.2$) and significant velocity deviations from local Keplerian ($\delta v$ up to $1.2c_{s}$) inside vortex region . It is expected that sources with modest inclination favors detection but requires long integration time with ALMA (more than 10h) to achieve the necessary signal-to-noise ratio.

\section{Summary and Future Prospects} \label{Sec:Conclusions}

In this work, we have systematically studied the dynamical consequence of thermal forcing by shadows cast to the outer protoplanetary disks. With a large survey of parameters, we have identified a diverse forms of substructures generated by shadows and studied their trends under different thermodynamic and viscosity prescriptions. Our results apply in regimes where the shadow is static or slowly-rotating (prograde), so that the corotation radius is further than regions of interest. The main findings of our studies are as follows.

1. Two-arm spirals with identical pattern speed as the shadow are fundamental substructures generated by weak thermal forcing ($\epsilon<=0.5$, $\sigma_{\phi}=0.079$, $\beta>1$) or high viscosity ($\alpha>10^{-3}$). They represent linear response to thermal forcing, and their pitch angle well agrees with standard density waves.
Both the density contrast (0.1-1$\%$ higher than background) and velocity disturbance up to 0.5$\%$ $v_K$) are small and scale with the strength of thermal forcing.

2. Disks with moderate thermal forcing are dominated by ring-like substructures. In this regime (parameter space in between crescent/vortex and spiral-dominant disks), the gas density contrast reaches 1-20$\%$ above the background. The rings are uniformly spaced ($\Delta r/H \sim 4H$) and exhibit pattern eccentricities on the order of $h/r$ or higher which rotates at the same rate of the shadow.

3. Crescents/vortices dominate disks under strong thermal forcing ($\epsilon>0.5$, $\beta\lesssim0.1$, $\sigma_{\phi}=0.236$) and low viscosity ($\alpha<=10^{-4}$). In this case, the density contrast is typically 10-50$\%$ higher than the average density at the same radius. The vortices in our simulations exhibit relatively large vorticity (ranging from 0.1 to 0.6, typically around 0.2) and significant velocity deviations from local Keplerian inside the vortex region (ranging from 0.4 to 1.2 $c_{s}$). Due to the chaotic nature (local turbulence level is 0.1 $c_{s}$) of the vortex-dominant disk, these structures are not uniformly spaced, with $\Delta r/H$ between 2 and 4.

4. Thermodynamics and viscosity significantly influence the formation of shadow-driven disk substructures. The dominant substructure transitions from spirals to rings and eventually to vortices as cooling timescales and/or viscosity decrease.

5. Owing to the simplicity of our problem setup, it is premature to definitely assess the observability of such shadow-driven substructures. We anticipate that the azimuthal brightness contrast in the sub-mm continuum to offer important constraints on the strength of the thermal forcing, while detecting in-plane kinematic signatures is likely challenging.

Through our suite of physically-motivated while highly simplified simulations, we highlight the importance on the dynamical impact of shadows or more generally, inhomogeneous stellar irradiation, on the gas dynamics of PPDs through thermal forcing. Given the fact that shadows are often observed in scattered light images of disks, our results call for proper consideration and incorporation of such effects for adequate modeling of such systems.

Our simulations can be considered as a starting point to understand the dynamical effects of shadows on PPDs, yet real systems are likely much more complex. This leaves several aspects to be considered and tested in the future. Proper characterizing disk thermodynamics is a pre-requisite to accurately model thermal forcing from  shadows, which requires better modeling of the shadow geometry, together with self-consistent radiation transport. Such modeling under typical disk parameters (that are likely nearly optically-thin) will likely reduce the azimuthal temperature contrast due to in-plane radiation transport.
Incorporation of dust dynamics is essential to obtain dust response to the shadow-driven substructures. Such simulations are expected to link the results with specific sources, as we are aware of efforts underway (Ziampras et al., in preparation). We have also assumed the shadows are cast to a full disk, whereas shadows are also observed in transition disks, and it is also pertinent to account for the interplay other physical mechanisms that cause disk substructures, with additional effect of self-shadowing.

Finally, all existing studies of shadow-driven disk dynamics are conducted in 2D in the disk plane, whereas the shadow-driven thermal forcing is also expected to also drive oscillations in the vertical direction (Liu \& Bai, in preparation). Future studies should incorporate 3D effects, which is essential to further assess the fidelity of 2D simulation results, and make more realistic observational predictions and comparisons.

{\bf Acknowledgements }
We thank Yanqin Wu and Shangjia Zhang for useful discussions, Pinghui Huang for helpful instructions on problem setup, and Alexandros Ziampras for constructive exchanges. This work is supported by National Science Foundation of China under grant No. 12233004, 12325304. We also acknowledge the China Center of Advanced Science and Technology for hosting the Protoplanetary Disk and Planet Formation Summer School in 2022 when this work was initiated. Numerical simulations are conducted in the Orion and Sirius clusters at Department of Astronomy, Tsinghua University and TianHe-1 (A) at National Supercomputer Center in Tianjin, China.

\appendix
\section{Transition State}\label{App:transition state}

\begin{figure*}[htbp]
    \centering
    \subfigure[]{
    \label{subFig:V_R_demo1}
    \includegraphics[width=0.2383\textwidth]{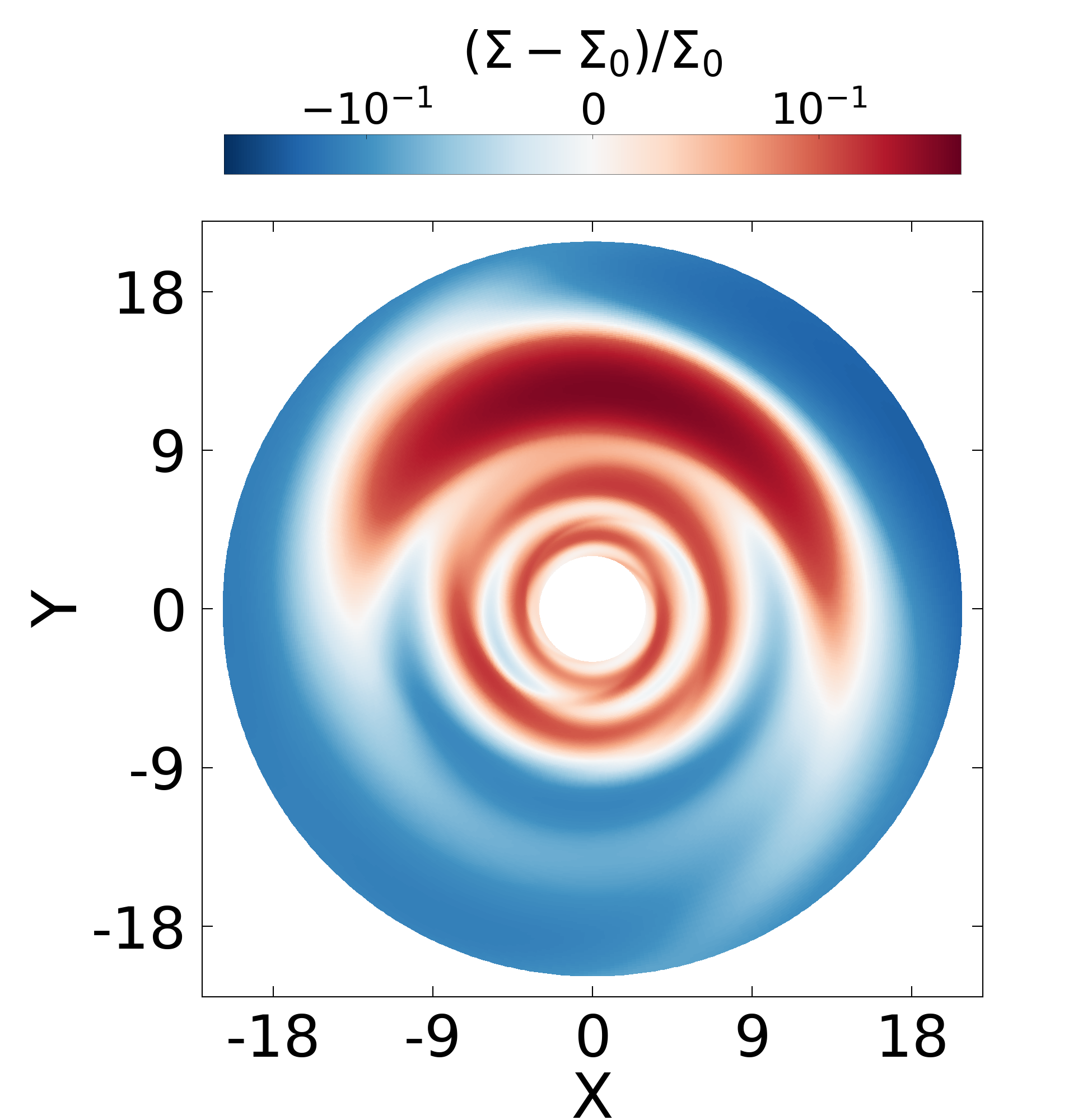}}
    \hspace{0.1cm}
    \subfigure[]{
    \label{subFig:V_R_demo2}
    \includegraphics[width=0.22\textwidth]{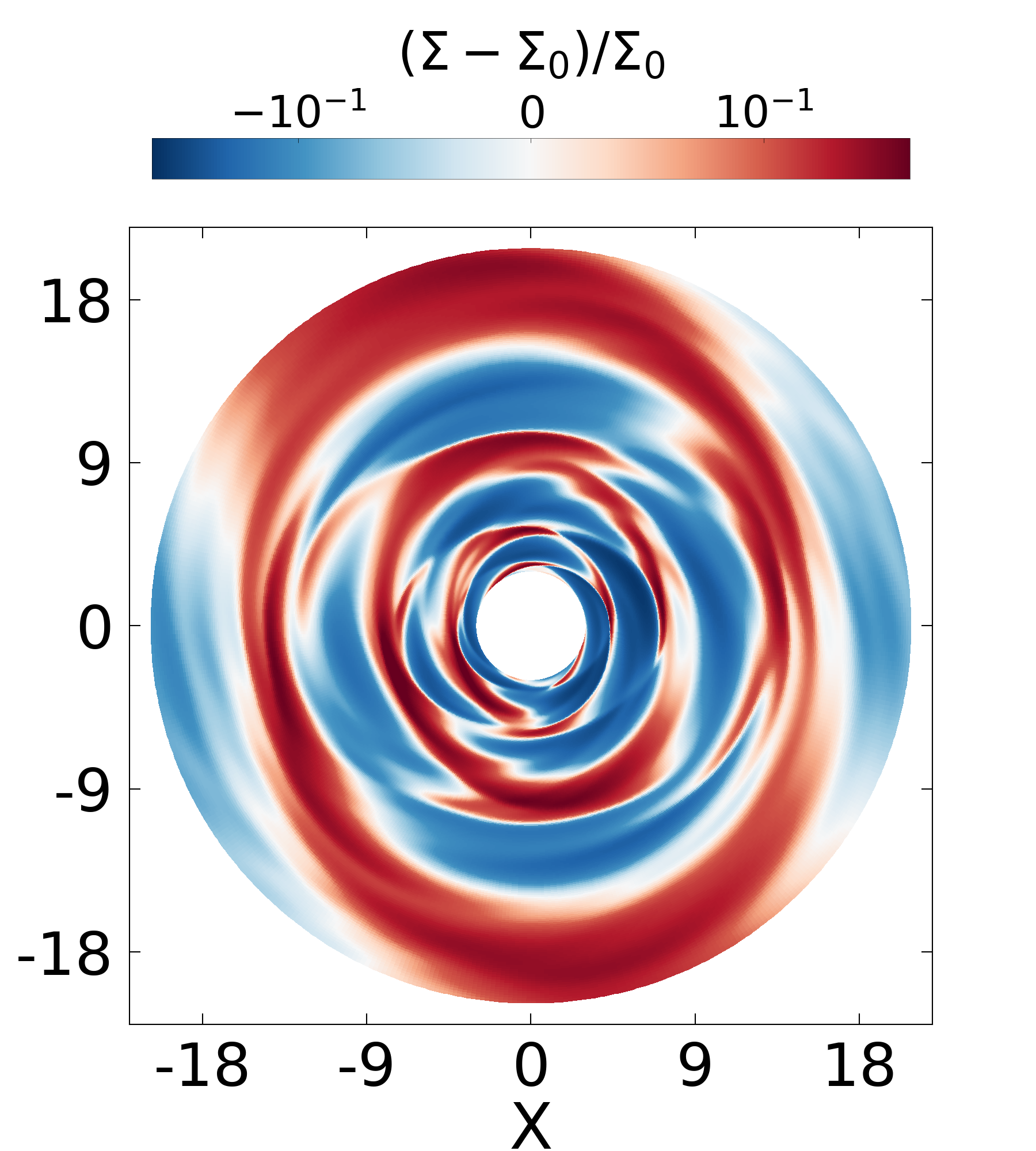}}
    \hspace{0.1cm}
    \subfigure[]{
    \label{subFig:V_R_demo3}
    \includegraphics[width=0.22\textwidth]{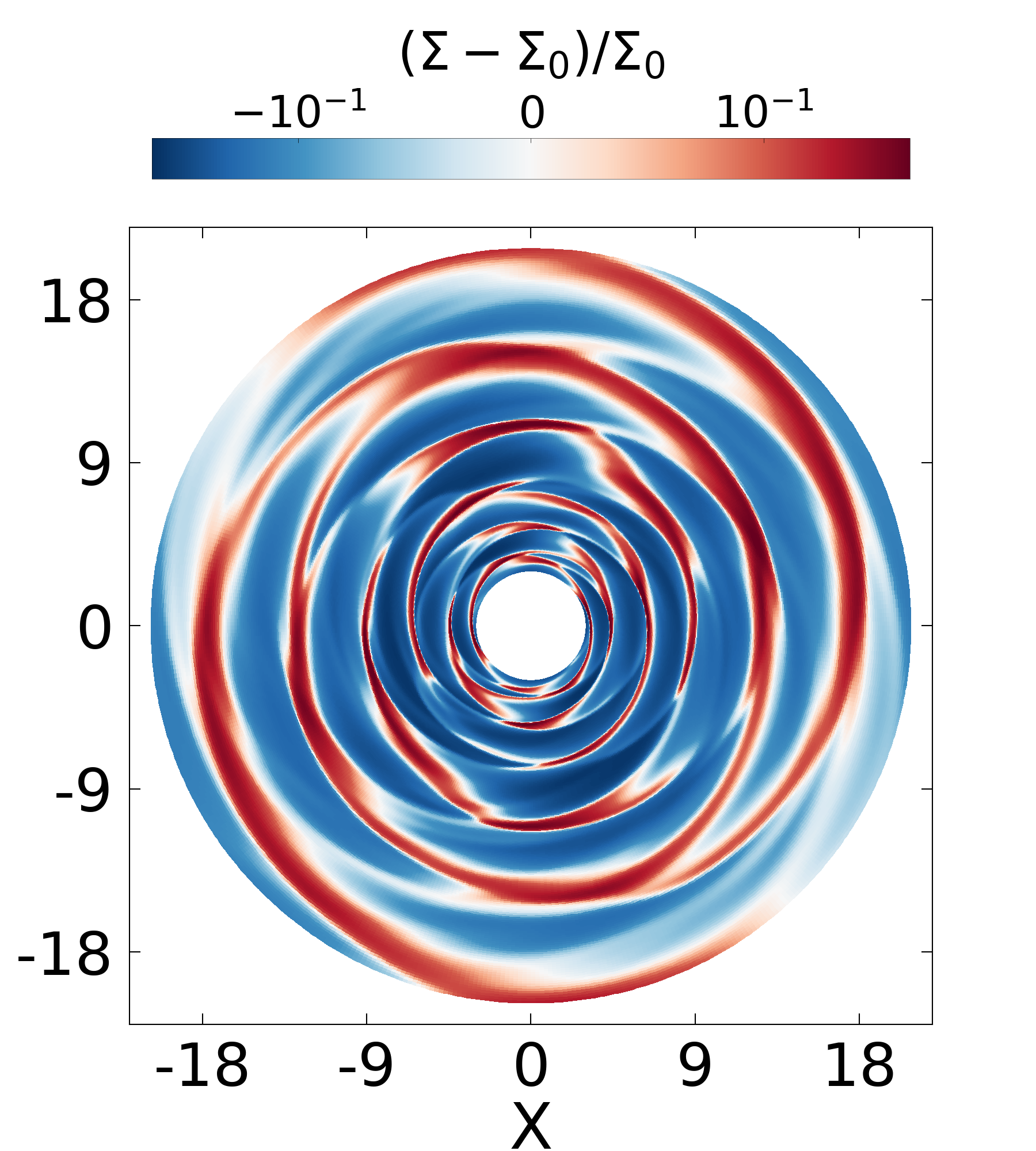}}
    \hspace{0.1cm}
    \subfigure[]{
    \label{subFig:V_R_demo4}
    \includegraphics[width=0.22\textwidth]{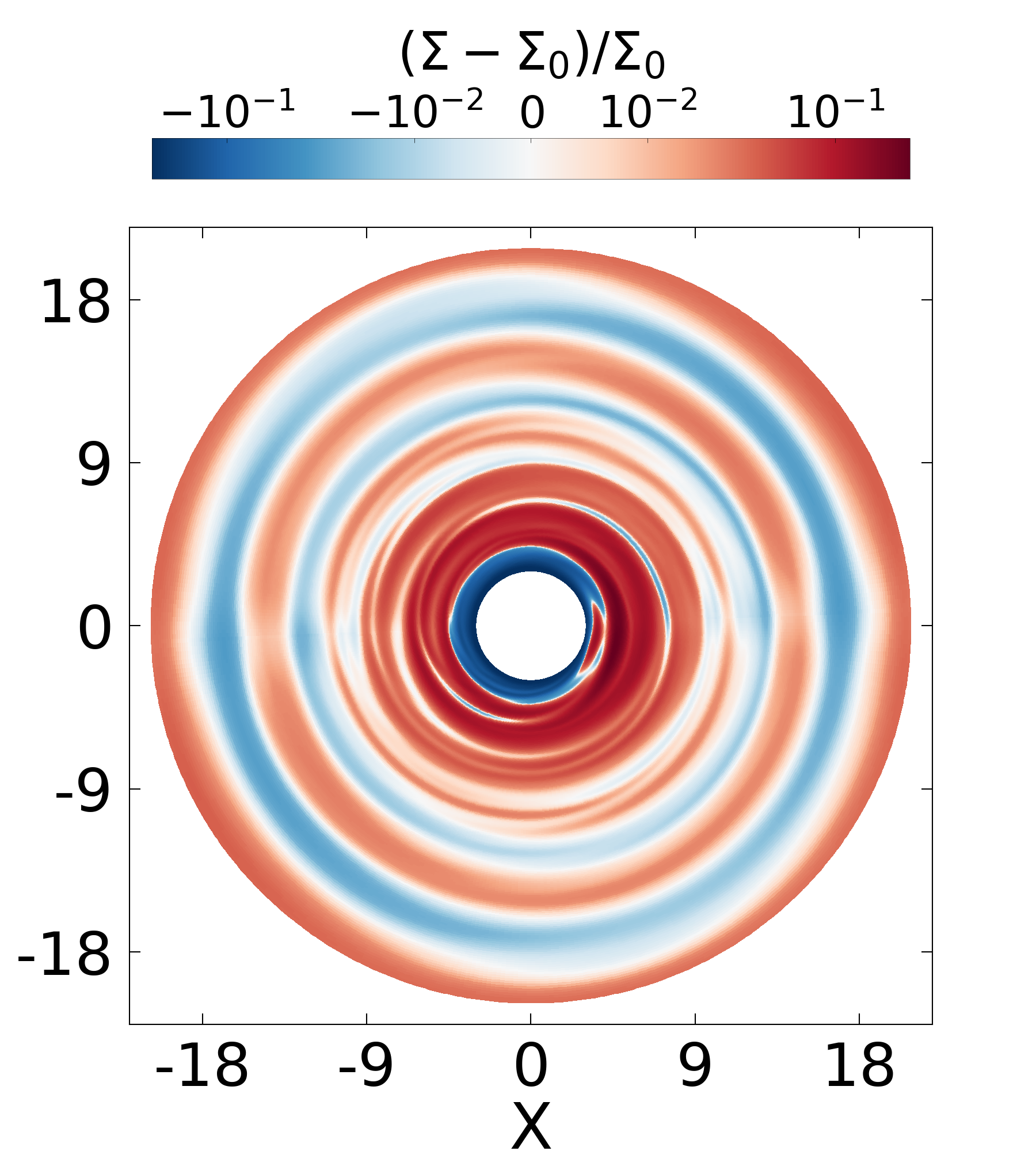}}
    \caption{The demonstration of vortex-ring transition situation. The parameters of these 4 plots are $\sigma_{\phi}=0.236, p=-0.5, \epsilon=0.5, \alpha=0.0001, \beta=1$; $\sigma_{\phi}=0.236, p=-0.5, \epsilon=0.8, \alpha=0.01, \beta=0.001$; $\sigma_{\phi}=0.236, p=-1.0, \epsilon=0.8, \alpha=0.01, \beta=0.001$; $\sigma_{\phi}=0.079, p=-0.5, \epsilon=0.8, \alpha=0, \beta=0.1$ respectively. All these cases can be found in Figure \ref{Fig:shadow_plot_int} and Figure \ref{Fig:shadow_plot_azw_and_pa} as vortex or ring blocks covered by red hatch lines.}
    \label{Fig:V_R_demo}
\end{figure*}

The vortex-ring transition represents the parameter regime where both the features of vortices/crescents and rings can be observed in the disk.  
Four examples of vortex-ring transition are illustrated in Figure \ref{Fig:V_R_demo}. They are recognized as vortex-ring transitions generally based on two reasons: rings and vortices/crescents are simultaneously present in the disk (Figure \ref{subFig:V_R_demo1}), or the basic morphology appears as rings but with significant asymmetry (Figure \ref{subFig:V_R_demo2}, \ref{subFig:V_R_demo3}, \ref{subFig:V_R_demo4}). In Figures \ref{Fig:shadow_plot_int} and \ref{Fig:shadow_plot_azw_and_pa}, the left side of vortex-ring transition cases depicts vortex-dominated disks, while the right side illustrates ring-dominated disks. Further decreases in $\beta$ or $\alpha$ lead to the disk being completely dominated by vortices/crescents.

\begin{figure*}[htbp]
    \centering
    \subfigure[]{
    \label{subFig:R_S_demo1}
    \includegraphics[width=0.2383\textwidth]{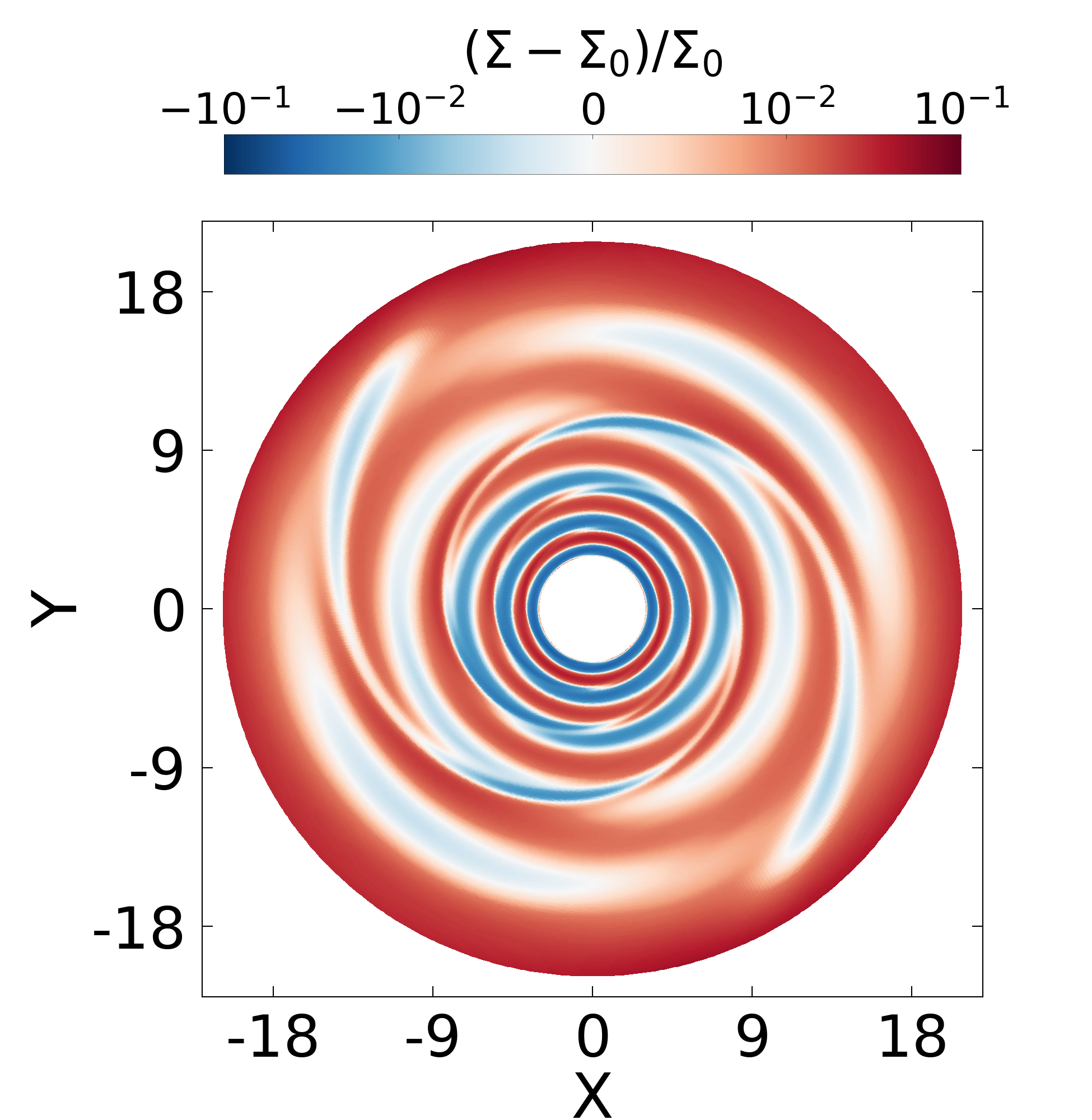}}
    \hspace{0.1cm}
    \subfigure[]{
    \label{subFig:R_S_demo2}
    \includegraphics[width=0.22\textwidth]{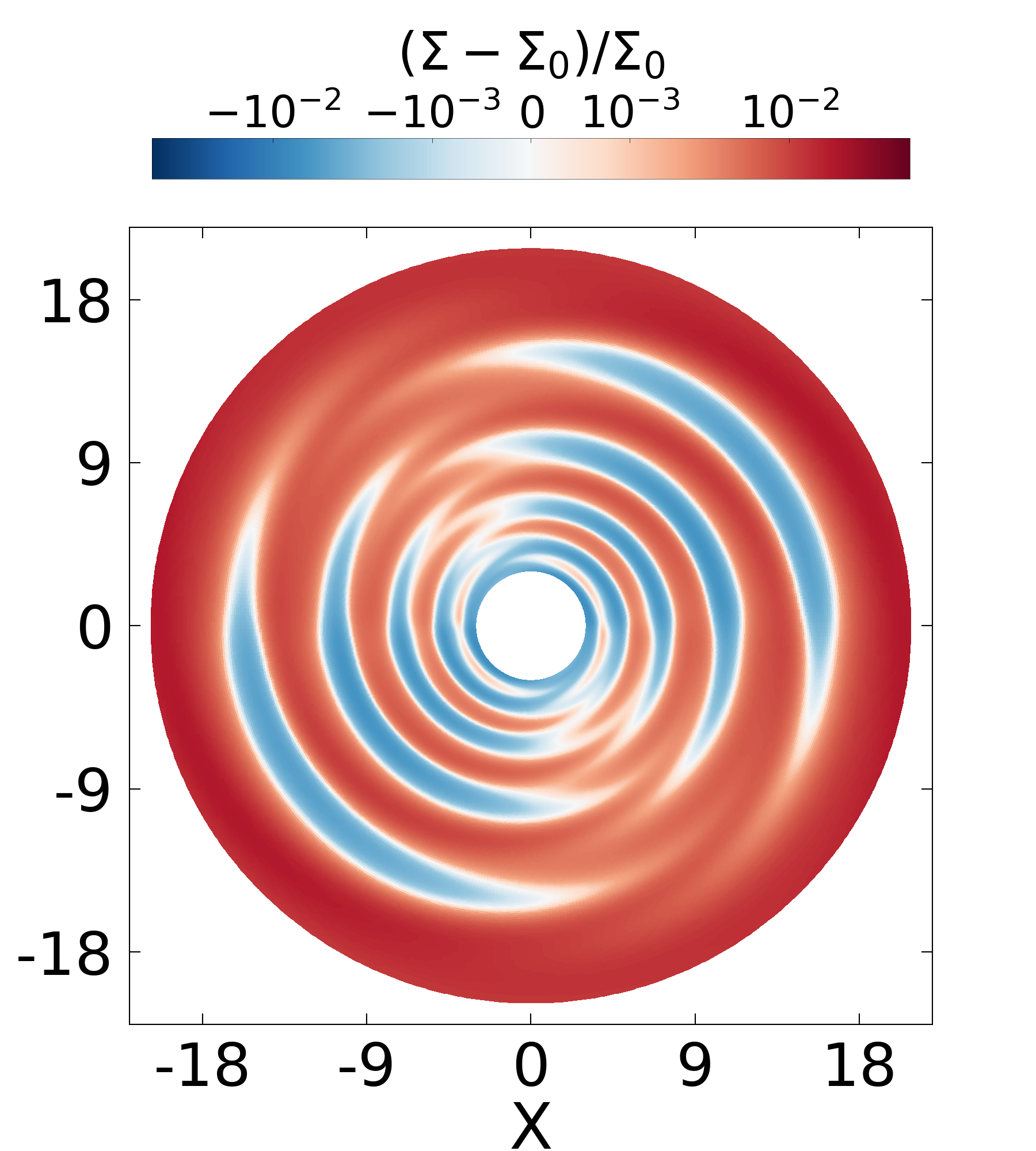}}
    \hspace{0.1cm}
    \subfigure[]{
    \label{subFig:R_S_demo3}
    \includegraphics[width=0.22\textwidth]{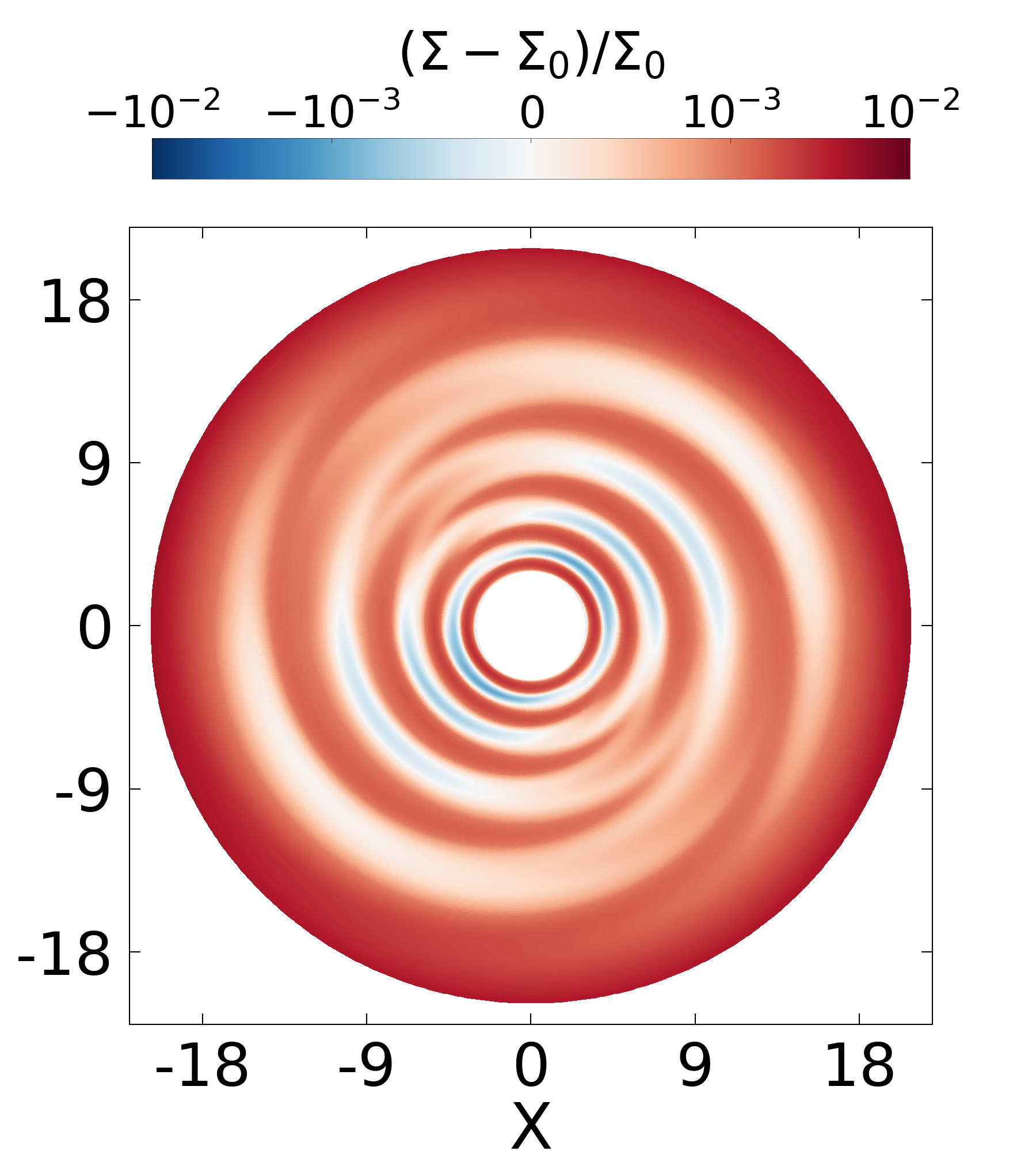}}
    \hspace{0.1cm}
    \subfigure[]{
    \label{subFig:R_S_demo4}
    \includegraphics[width=0.22\textwidth]{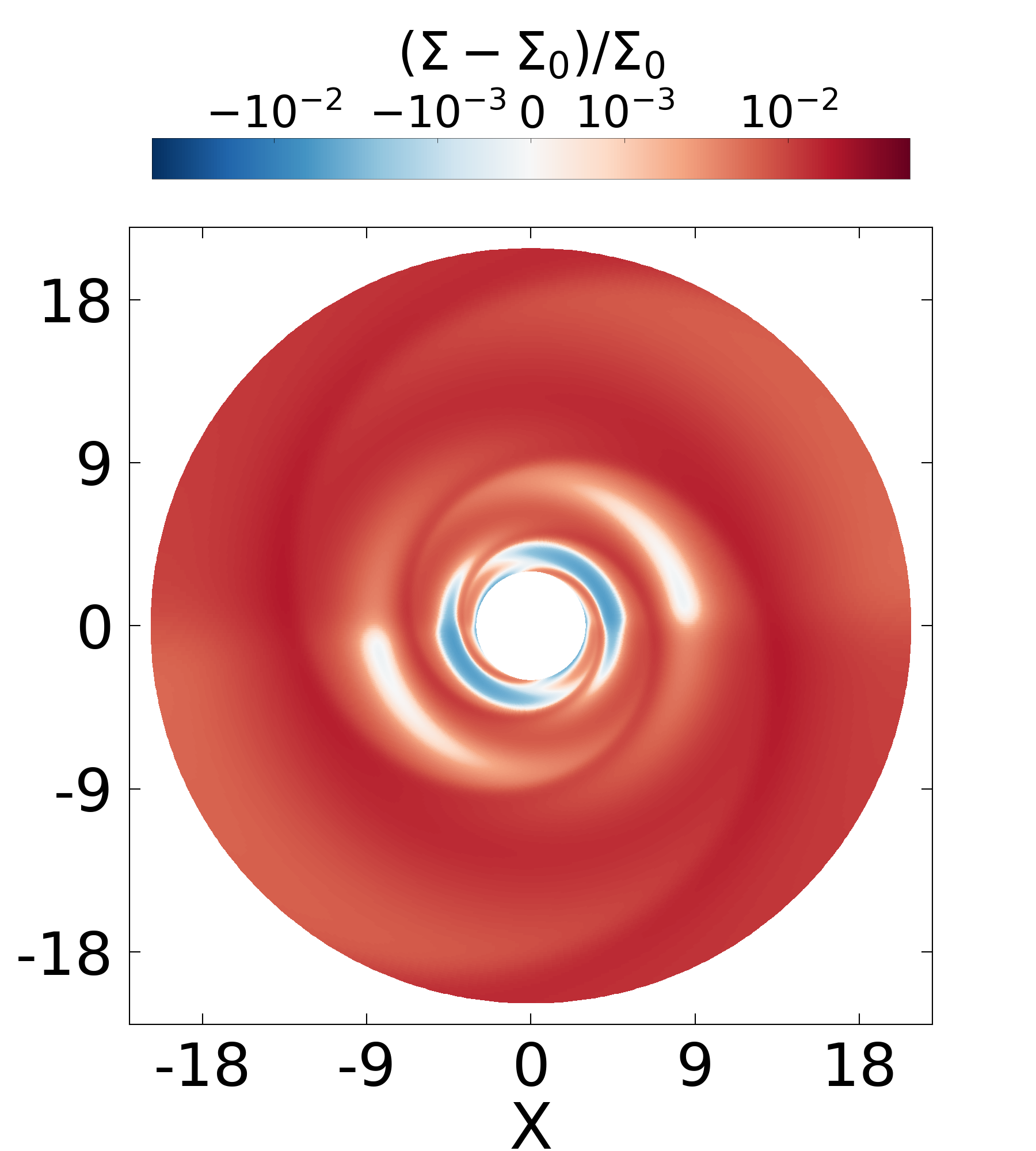}}
    \caption{The demonstration of ring-spiral transition situation. The parameters of these 4 plots are $\sigma_{\phi}=0.236, p=-0.5, \epsilon=0.8, \alpha=0, \beta=10$; $\sigma_{\phi}=0.079, p=-1.0, \epsilon=0.8, \alpha=0.001, \beta=1$; $\sigma_{\phi}=0.079, p=-0.5, \epsilon=0.5, \alpha=0, \beta=10$; $\sigma_{\phi}=0.079, p=-0.5, \epsilon=0.5, \alpha=0.001, \beta=1$ respectively. All these cases can be found in Figure \ref{Fig:shadow_plot_int} and Figure \ref{Fig:shadow_plot_azw_and_pa} as ring or spiral blocks covered by blue hatch lines.}
    \label{Fig:R_S_demo}
\end{figure*}

The ring-spiral transition represents the parameter regime where both the features of rings and spirals can be identified in the disk.
%This implies that the disk becomes dominated by spirals as $\beta$ or $\alpha$ increases. 
Four examples of ring-spiral transitions are shown in Figure \ref{Fig:R_S_demo}. They either exhibit regularly broken rings (Figure \ref{subFig:R_S_demo2} and \ref{subFig:R_S_demo3}) or clearly display both rings and spirals within the same disks (Figure \ref{subFig:R_S_demo1} and \ref{subFig:R_S_demo4}). These transition regions lie between ring-dominated disks and spiral-dominated disks in Figure \ref{Fig:shadow_plot_int} and \ref{Fig:shadow_plot_azw_and_pa}. The disk becomes dominated by spirals as $\beta$ or $\alpha$ increases.

From the transition states shown in Figure \ref{Fig:V_R_demo} and \ref{Fig:R_S_demo}, we can verify that rings exhibit characteristics of both vortices/crescents and spirals, as discussed in Section \ref{subSec:sts_ring}. Slightly excessive thermal forcing, relative to ring-dominant disks, can hamper reconnection (Figure \ref{subFig:V_R_demo3}) mentioned in Section \ref{subSec:nonlinear regime}, leading the disk into a vortex-ring transition state with strongly asymmetric rings (Figure \ref{subFig:V_R_demo2}) or crescents with large aspect ratios (Figure \ref{subFig:V_R_demo1}). Conversely, with weak thermal forcing, the breaking of two-armed spirals is partial (Figure \ref{subFig:R_S_demo3}), placing the disk into a ring-spiral transition state.

\section{Simulation statistics} \label{App:detail_sts}

\begin{figure*}[htbp]
    \centering
    \includegraphics[width=0.85\textwidth]{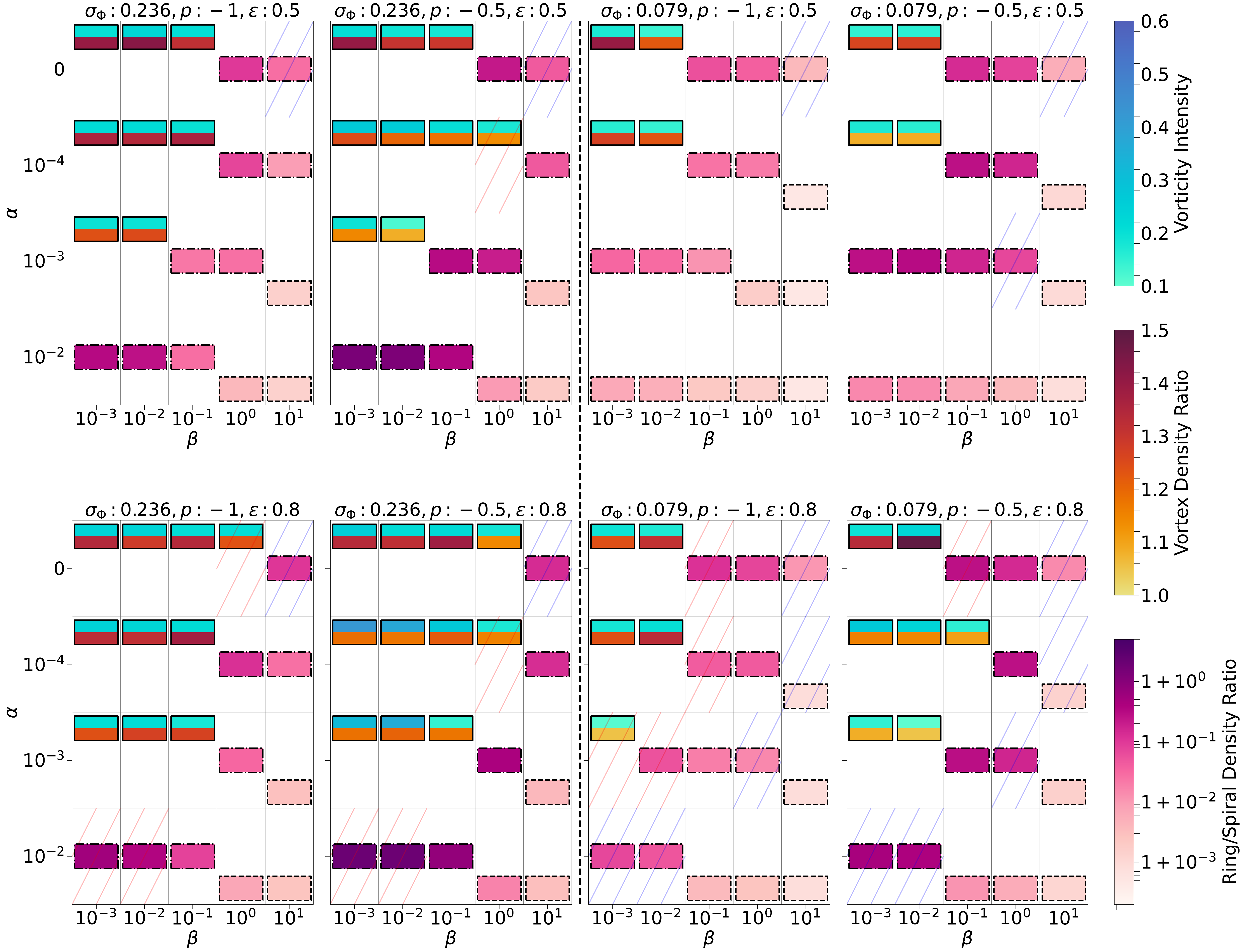}
    \caption{Statistics of substructures’ vorticity and density ratio for 160 simulations. The structure of the figure is described in Appendix \ref{App:detail_sts} and is similar to Figure \ref{Fig:shadow_plot_int_sim}.}
    \label{Fig:shadow_plot_int}
\end{figure*}

\begin{figure*}[htbp]
    \centering
    \includegraphics[width=0.85\textwidth]{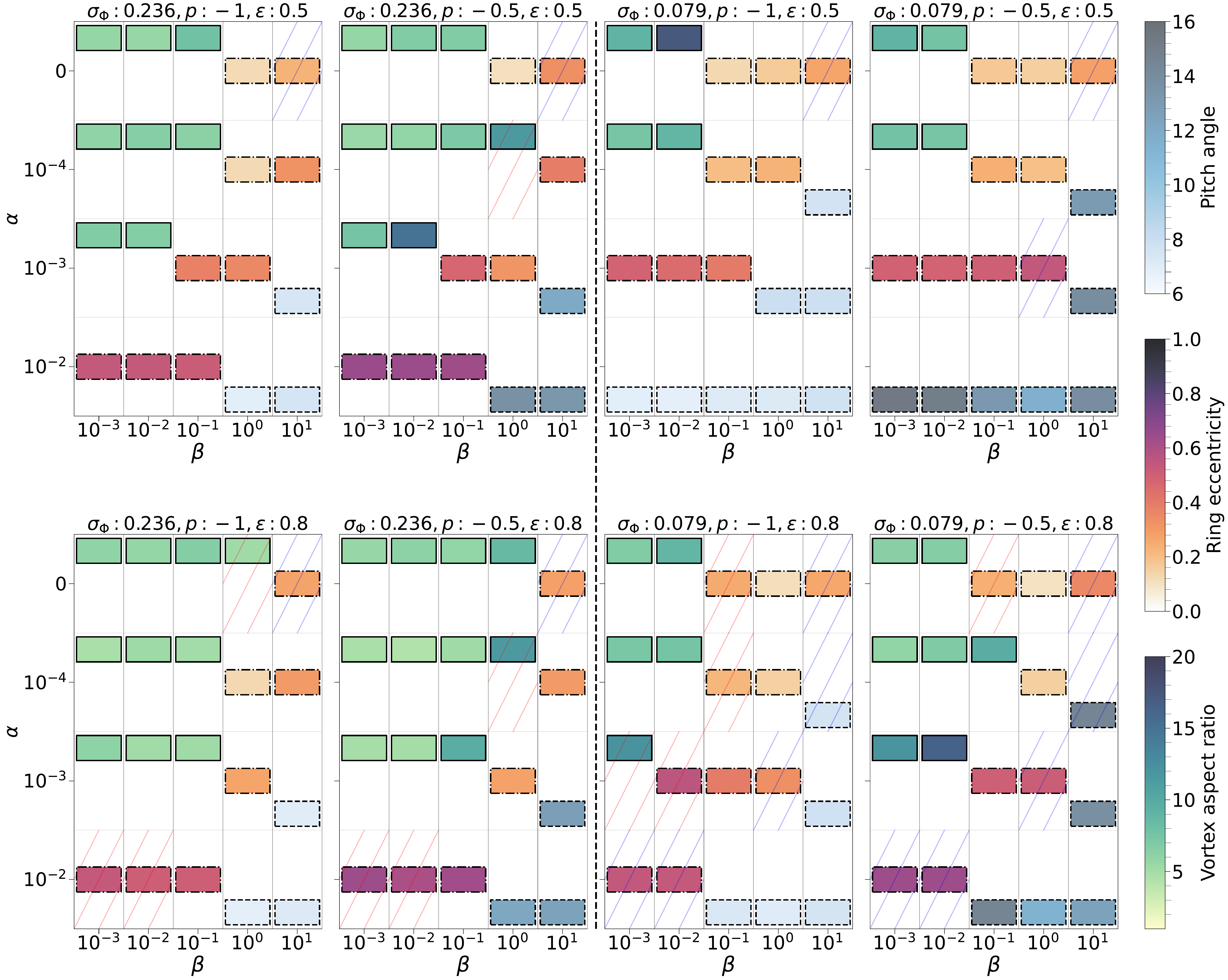}
    \caption{Statistics of vortices’ aspect ratio, rings' eccentricity and spirals' pitch angles for 160 simulations. The structure of the figure is same as Fig\ref{Fig:shadow_plot_int}.}
    \label{Fig:shadow_plot_azw_and_pa}
\end{figure*}

The detailed statistical plot of vorticity and density contrast (Figure \ref{Fig:shadow_plot_int}), along with other parameters (Figure \ref{Fig:shadow_plot_azw_and_pa}) of substructures, is presented here. These two figures share the same structures. Each of these figures is divided into two sections by a dashed line, representing shadow ranges of 45 degrees ($\sigma_{\phi}=0.236$) and 15 degrees ($\sigma_{\phi}=0.079$), respectively. In the left column, disks with a temperature slope of $-1$ are shown, while the right column represents disks with a temperature slope of $-0.5$. Each row, from top to bottom, corresponds to
% perturbation strengths 
shadow amplitudes of $\epsilon=0.5$ and $\epsilon=0.8$. The $x$-axis of the subfigures represents $\beta$, while the $y$-axis represents $\alpha$. Within each $\beta$-$\alpha$ section, there are three rows indicating the dominant structures in the disk: vortices/crescents, rings, and spirals, each represented by different types of squares, colored by the relevant physical properties as indicated in the color bars. The figure also includes red and blue line shaded areas, indicating disks undergoing transitions from vortex-ring and ring-spiral phases, respectively. We note that the inviscid ($\alpha=0$) simulations maintain the same temperature gradient with a slope of $-1$ (indicating that the $p$ value shown in the title of each subfigure only applies to viscid runs) and vary the density gradient of $d=-0.5$ and $d=-1$ in the left and right columns, respectively, which help us exclude the influence of density gradient.

\section{One-sided shadow test}\label{App:one shadow}

\begin{figure*}[htbp]
    \centering
    \subfigure[]{
    \label{subFig:spiral_one}
    \includegraphics[width=0.325\textwidth]{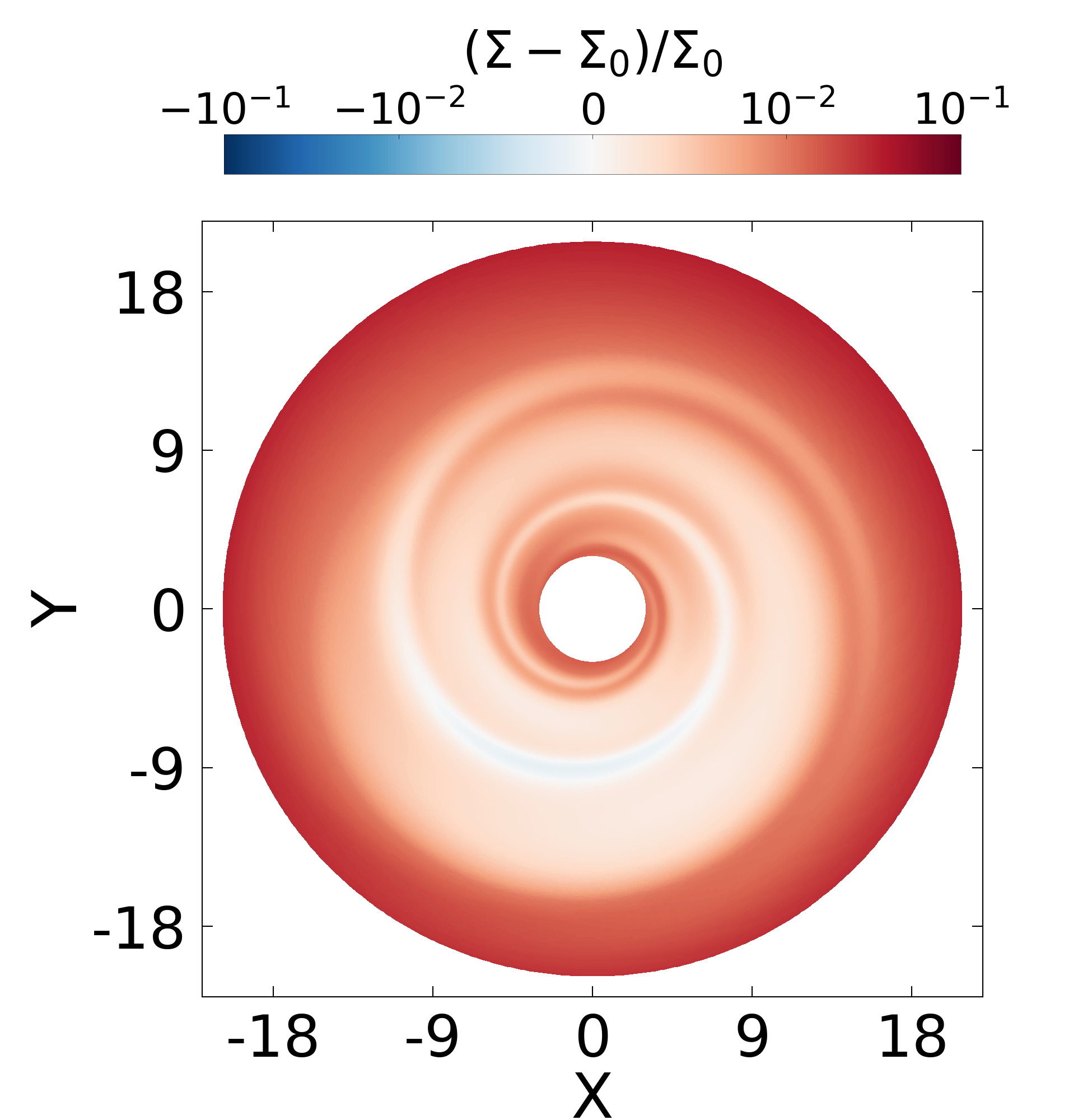}}
    \hspace{0.1cm}
    \subfigure[]{
    \label{subFig:ring_one}
    \includegraphics[width=0.3\textwidth]{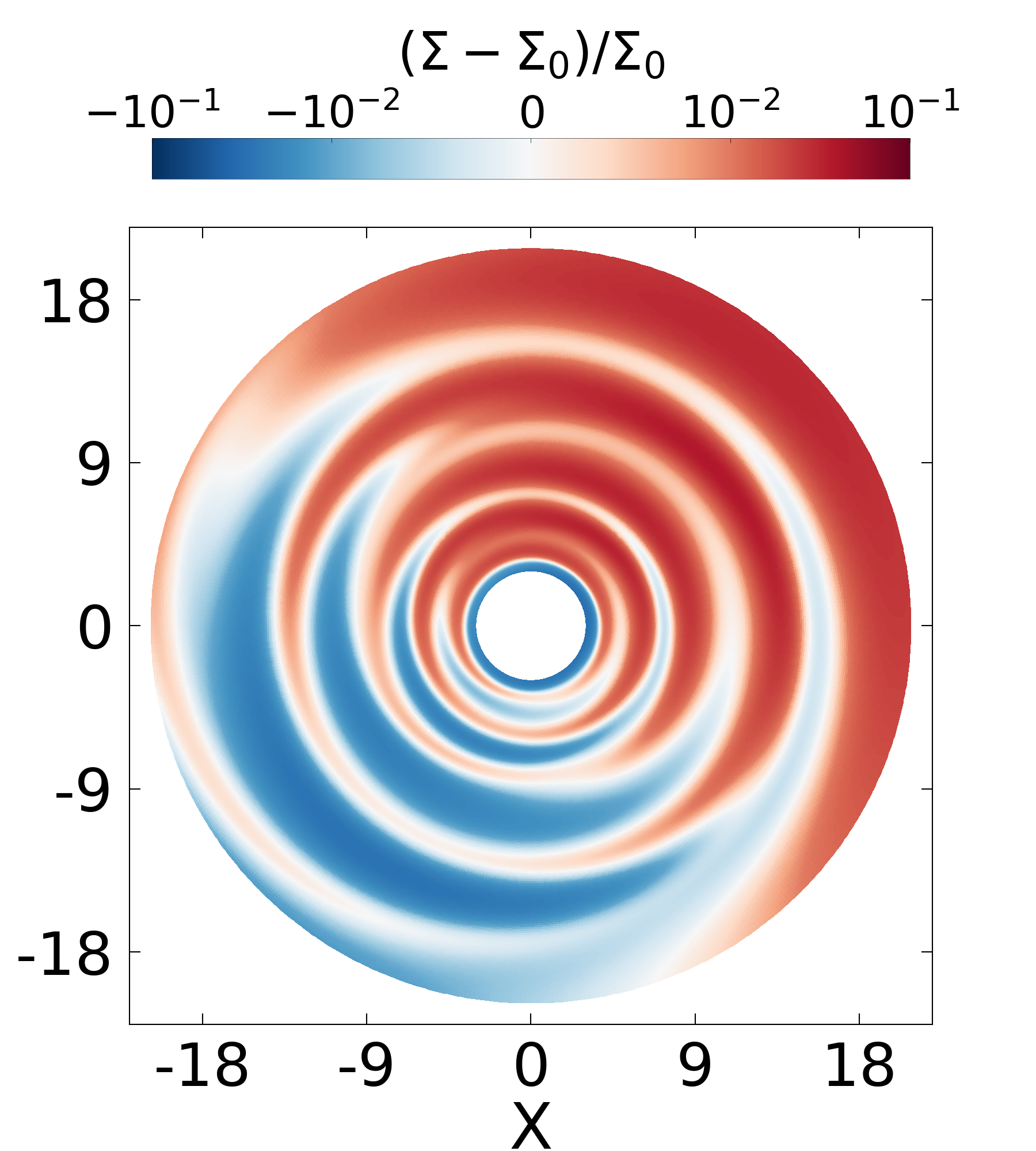}}
    \hspace{0.1cm}
    \subfigure[]{
    \label{subFig:vortex_one}
    \includegraphics[width=0.3\textwidth]{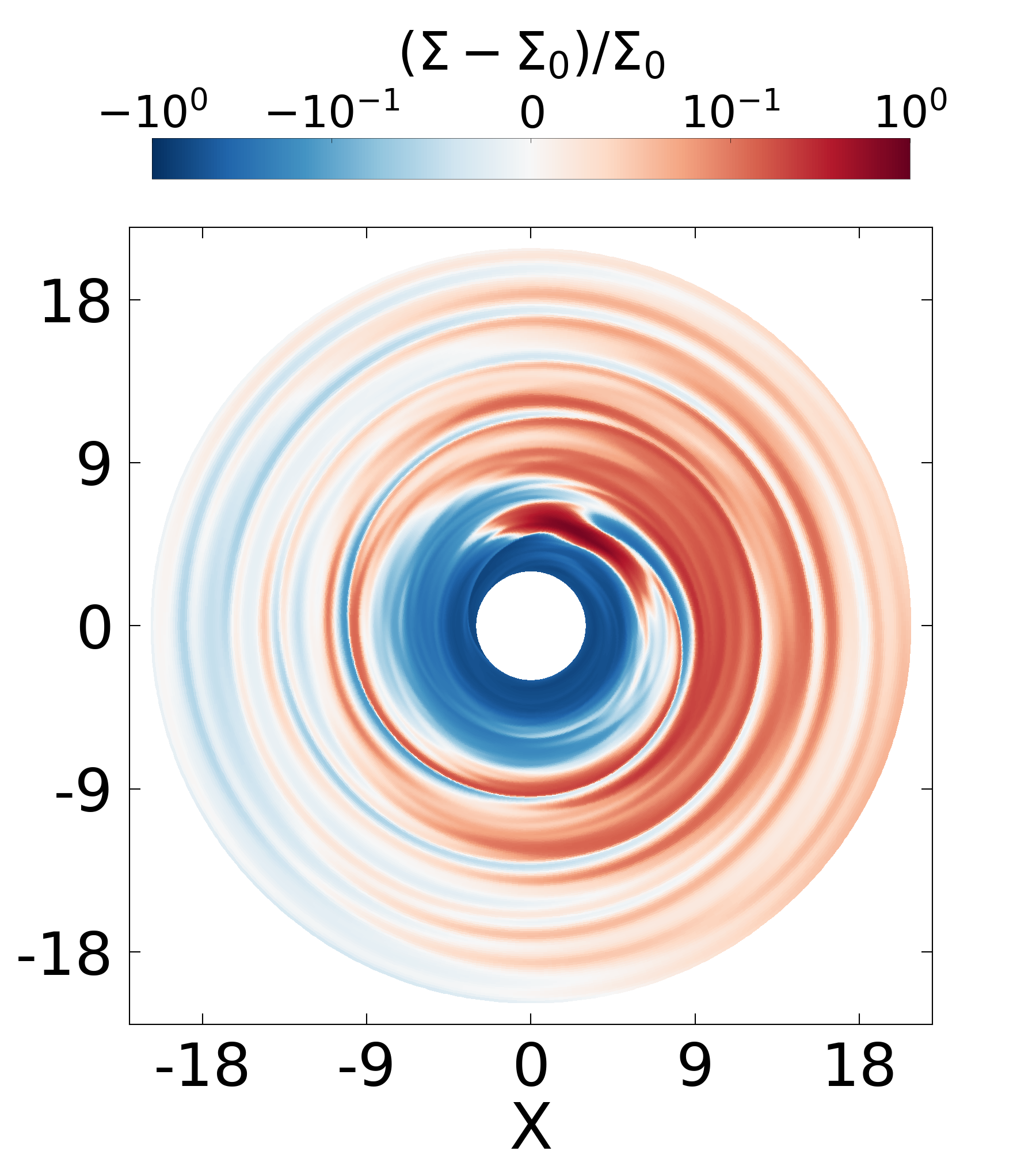}}
    \caption{Normalized density perturbation for one-sided shadow. Left panel: Spiral forming disk with parameter taken to be same as run NL-hm-S-NR (Figure \ref{Fig:spiral}). Middle panel: Ring forming disk with parameter taken to be same as run NL-hm-R-NR (Figure \ref{Fig:ring}). Right panel: Vortex forming disk with parameter taken to be same as run NL-hm-V-NR (Figure \ref{Fig:vortex}).}
    \label{Fig:one shadow}
\end{figure*}

In this Appendix, we briefly examine how the morphology and form of substructures can be affected by the morphology of the shadow region. As an experiment,
we performed simulations with only the right side of the shadow shown in Figure \ref{Fig:temperature} present, and the target temperature is taken as
\begin{equation}\label{Equ:one_shadow_temp}
T_{\rm tar} (r,\phi)= T_{\rm init}(r)\left( 1-\epsilon e^{-\frac{\phi^{2}}{2\sigma_{\phi}^{2}}}\right).
\end{equation}
The remaining parameters for the disk and shadow are the same as those in the representative simulations (NL-hm runs). For detailed parameter settings for the NL-hm runs, please refer to Table \ref{Tab:table-1}.
It can be seen from Figure \ref{Fig:one shadow} that the types of dominant substructures have not changed compare with NL-hm runs. The dominant spiral now has $m=1$, and the rings become asymmetric (with $m=1$, as opposed to eccentric with $m=2$), while crescents are generated as usual. These outcomes similarly follow the formation process described in Section \ref{subSec:nonlinear regime}. These simulations illustrate that besides a morphological change from $m=2$ to $m=1$, the general trends of shadow-driven substructures are not sensitive to shadow prescriptions.

%%%%%%%%%%%%%%%%%%%%%%%%%%%%%%%%%%%%%%%%%%%%%%%%%%%%%%%
%%% Reference section. 

\bibliography{shadow}{}
\bibliographystyle{aasjournal}

\end{document}